\documentclass[aps,
prx,
footinbib,
floats,
twocolumn,
showpacs,
superscriptaddress,
]{revtex4-1}

\usepackage{natbib}
\usepackage[utf8]{inputenc}
\usepackage{amsmath, amsthm, amssymb}
\usepackage{enumitem}
\usepackage{graphics,graphicx}
\usepackage{epsfig}
\usepackage{epstopdf}
\usepackage{times}
\usepackage{dcolumn}
\usepackage{url}
\usepackage{color}
\usepackage{stackengine}
\usepackage{tabularx}
\usepackage[all]{xy}
\usepackage{mdwlist}
\usepackage{xr}
\usepackage{float}

\begin{document}

\renewcommand*\thesection{\arabic{section}}
\newcommand{\beq}{\begin{equation}}
\newcommand{\eeq}{\end{equation}}
\newcommand{\sss}{\scriptscriptstyle}

\title{Antagonistic Structural Patterns in Complex Networks}

\author{Mar\'ia J.~Palazzi}
\affiliation{Internet Interdisciplinary Institute (IN3), Universitat Oberta de Catalunya, Barcelona, Catalonia, Spain}

\author{Javier Borge-Holthoefer}
\affiliation{Internet Interdisciplinary Institute (IN3), Universitat Oberta de Catalunya, Barcelona, Catalonia, Spain}

\author{Claudio J.~Tessone}
\affiliation{URPP Social Networks, Universit\"at Z\"urich, Switzerland}

\author{Albert Sol\'e-Ribalta}
\affiliation{Internet Interdisciplinary Institute (IN3), Universitat Oberta de Catalunya, Barcelona, Catalonia, Spain}

\begin{abstract}
Identifying and explaining the structure of complex networks at different scales has become an important problem across disciplines. At the mesoscale, modular architecture has attracted most of the attention. At the macroscale, other arrangements --e.g. nestedness or core-periphery-- have been studied in parallel, but to a much lesser extent. However, empirical evidence increasingly suggests that characterizing a network with a unique pattern typology may be too simplistic, since a system can integrate properties from distinct organizations at different scales. Here, we explore the relationship between some of those organizational patterns: two at the mesoscale (modularity and in-block nestedness); and one at the macroscale (nestedness). We analytically show that nestedness can be used to provide approximate bounds for modularity, with exact results in an idealized scenario. Specifically, we show that nestedness and modularity are antagonistic.
Furthermore, we evince that in-block nestedness provides a parsimonious transition between nested and modular networks, taking properties of both. Far from a mere theoretical exercise, understanding the boundaries that discriminate each architecture is fundamental, to the extent modularity and nestedness are known to place heavy constraints on the stability of several dynamical processes, specially in ecology.
\end{abstract}


\maketitle


The detection and identification of emergent structural patterns has been a main focus in the development of modern network theory. Such interest is not surprising, because these arrangements lie at the core of the discipline as one of the keys to the origins --which are the assembly rules that led to an observed pattern?-- and dynamics --how is the system's activity constrained by the structure?-- of a network. In addition to these essential questions, the identification of structural signatures is a difficult task {\it per se}, which explains as well why so much attention has been put on the technical problem.

Undoubtedly, in this context, modularity \cite{newman2004finding,newman2006modularity} stands out: the organization of a network as a set of cohesive subgroups has, by far, concentrated most of the efforts \cite{duch2005community,blondel2008fast,newman2004analysis,leicht2008community,barber2007modularity}. Modular architecture is widespread 
\cite{zachary1977information,guimera05,eriksen2003modularity,adamic2005political,fortunato2010community} and responds to the intuition that similar elements in a complex system tend to flock together. 
However, there are other architectural principles beyond community structure which may play more important roles. In ecology, for instance, scholars have faced the need to define and quantify other patterns, demonstrated to be more relevant in certain scenarios. This is the case of nestedness \cite{atmar1993measureorder,bascompte2003nested}, 
a concept that has been crucial to understand the stability and diversity of ecological systems \cite{Thebault2010,bastolla2009architecture}. Other, more intricate, possibilities have also been explored \cite{lewinsohn2006structure,almeida2007nestedness}, like core-periphery structures \cite{borgatti2000models,rombach2017core} --and its extension to the mesoscale \cite{kojaku2017finding}-- are good examples of comparatively less studied architectures.

In these settings, the accent has been mainly placed on designing heuristics and improving algorithms \cite{rombach2017core,fortunato2010community,lin2018nestedness}; understanding the dynamical constraints that those patterns impose \cite{arenas2006synchronization,allesina2012stability}; or describing plausible microscopic rules that make those patterns emerge \cite{leung2016conflicting,Thebault2010,konig2014nestedness,verma2016emergence}. However, we have very limited knowledge on how different structural signatures may be intertwined, or how --if ever-- they affect and limit each other. Indeed, we have examples in which two or more structural features (say, nestedness and modularity) have been jointly considered \cite{olesen2007modularity,fortuna2010nestedness,flores2011statistical,borge2017emergence}. But such consideration overlooked to what extent the inherent constraints of one pattern limit --or boost-- the presence of the other. Furthermore, there is extensive evidence that modular and nested architectures play a critical role in relation to the stability of the dynamics of ecological systems \cite{bastolla2009architecture,Thebault2010,stouffer2011compartmentalization,allesina15}, economics \cite{bustos2012dynamics,Saavedra2011} and social sciences \cite{borge2017emergence}. Thus, understanding the possibility of coexistence of these structural patterns may shed light on the dynamical trade-offs that either arrangement can facilitate.

In \cite{borge2017emergence}, the authors observe that modularity and nestedness exhibit an anti-correlated behavior, suggesting that, at least in empirical data, these arrangements hardly coexist. In this work, we depart from this shallow evidence to first explore experimentally, and then analytically, the structural relationship between three structural patterns: nestedness on the macroscopic side,  modularity and in-block nestedness at the mesoscale. We analytically characterize these measures in an idealized family of networks, which allows us to precisely derive to what extent the macroscale organization places strict bounds to the emergence of mesoscale patterns. Eventually, we provide soft bound estimations for less restricted scenarios, {\it i.e.}~
real networks.


In a perfectly nested network, the set of neighbors of lower degree nodes are a subset of those  with larger degree \cite{cite-foot}. Such network is typically represented by a presence-absence matrix $a_{ij}$,   and the degree of nestedness $\mathcal{{N}}$, can be formally defined as \cite{sole2018revealing}
\begin{equation}
\mathcal{{N}} = \frac{2}{N_{T}} \left \{  \sum_{ij}^{N_{T}} \left[  \frac{\mathit{O}_{ij}- \langle \mathit{O}_{ij}\rangle}{k_{j}(N_{T} - 1)} \Theta( k_{i} -k_{j} )\right]  \right \},
\label{eq_nest}
\end{equation}
where $O_{ij} = \sum_{k}{a_{ik}a_{jk}}$ quantifies the overlap between nodes $i$ and $j$; $k_i$ corresponds to the degree of node $i$; and $\Theta(\cdot)$ is the Heaviside step function, that ensures that $O_{ij}$ has a positive contribution when $k_{i} \ge k_{j}$. Additionally, $O_{ij}$ is conveniently corrected by a null model that discounts the expected overlap if links where drawn randomly, $\langle \mathit{O}_{ij}\rangle$. $N_{T}$ is the size of the network. 


A modular structure is a rather ubiquitous mesoscale structural organization in which nodes are organized forming groups, i.e.~devoting many links to nodes in the same group, and fewer links towards nodes outside. One of the most popular methods to identify communities is through the maximization of the modularity $Q$ \cite{newman2004finding}. The original equation can be rewritten as 
\begin{equation}
 Q=\sum_{c=1}^{B}\left[ \frac{l_c}{L}- \left(\frac{d_c}{2L}\right)^2 \right], \label{eq:mod}
\end{equation}
where  $B$ is the number of communities, $L$ is the total number of links in the network, $l_c$ is the total number of links in community $c$, and $d_c$ is the sum of the degrees of all nodes in such community.

The possibility of a combined nested-modular organization has been debated in different contexts \cite{flores2013multi,beckett2013coevolutionary,sole2018revealing}. One conceivable form of coexistence is in terms of hybrid structures, as described by Lewinsohn {\it et al}. in \cite{lewinsohn2006structure}. The general layout of these networks is modular, but interactions within each module (or block) are expected to be nested. In contrast, modularity makes no assumption on the internal organization of communities. Worth highlighting, this hybrid structure reframes nestedness, originally a macroscale feature, to the mesoscopic level --it can now be interpreted as an in-block nested structure with $B=1$. 
The degree of in-block nestedness of a network $\mathcal{I}$ \cite{sole2018revealing} can be computed as 
\begin{equation}
\mathcal{I} = \frac{2}{N_{T}} \left \{  \sum_{i,j}^{N_{T}} \left[  \frac{\mathit{O}_{ij}- \langle \mathit{O}_{ij}\rangle}{k_{j}(C_{i} - 1)} \Theta( k_{i} -k_{j} ) \delta (\alpha _{i}, \alpha _{j}) \right] \right \}.
\label{eq_inb}
\end{equation}
Here, $\alpha_i$ represents the community node $i$ belongs to, and $C_{i}$ its size.  $\delta(\alpha_i,\alpha_j)$ corresponds to the Kronecker delta, equal to one only if nodes $i$ and $j$ belong to the same community. 


In order to experimentally assess the dependence between the different presented structural measures, we rely on a probabilistic network generation model \cite{sole2018revealing}. This model is able to generate structures (with and without noise) that smoothly interpolate between the different structural patterns of interest by recourse of 4 parameters: 
the number of modules $B$, the fraction of inter-modular links $\mu$ (inter-block noise), the fraction of links outside the perfect nested structure $p$ (intra-block noise), and the shape of the nested structure $\xi$ (cf.~Supplemental Material). 


\begin{figure*}
\centering
\def\tabularxcolumn#1{m{#1}}
\begin{tabularx}{\linewidth}{@{}cXX@{}}
\def\stackalignment{l}
\topinset{\bfseries(a)}{\includegraphics[width=0.5\columnwidth]{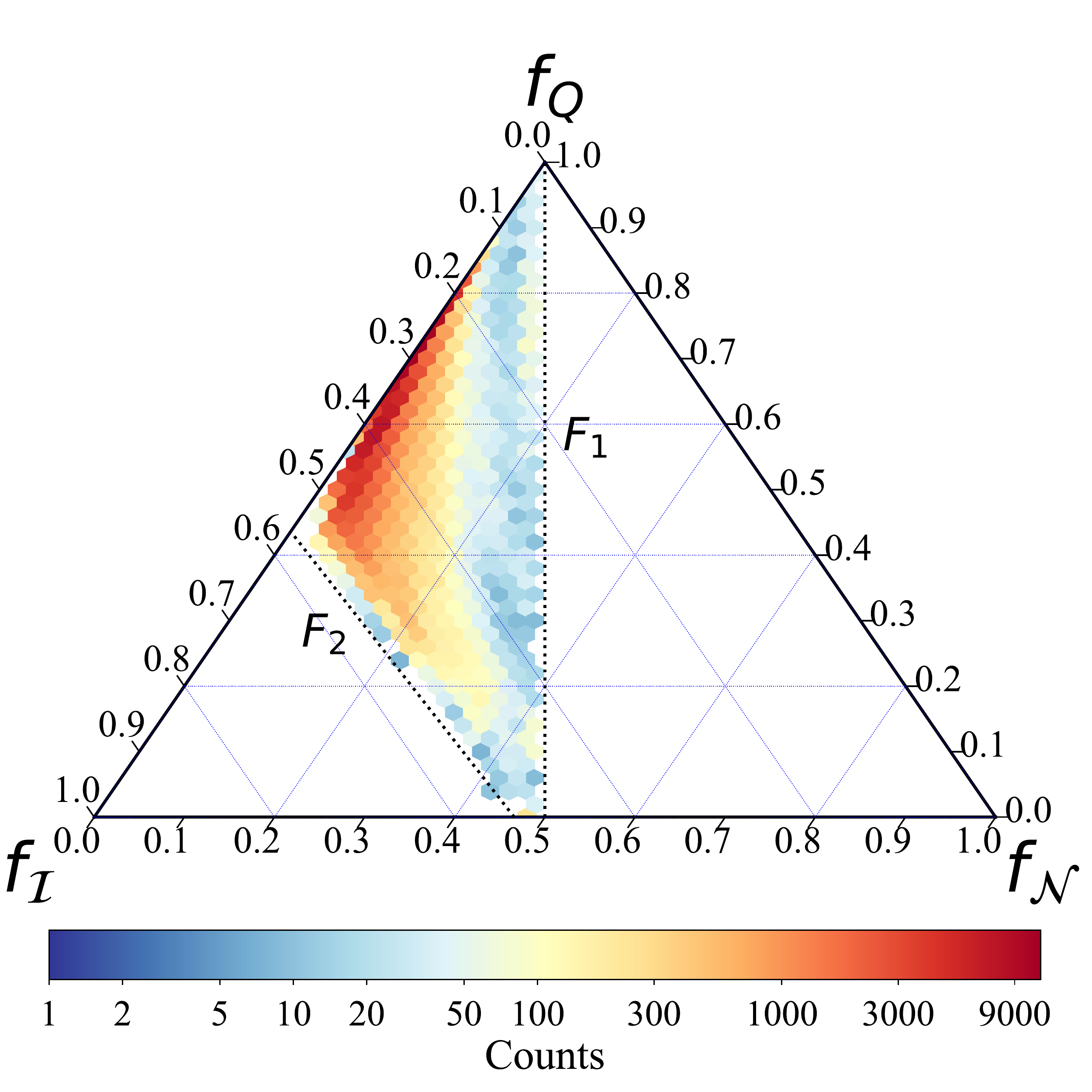}}{0.2in}{0.05in}
\topinset{\bfseries(b)}{\includegraphics[width=0.5\columnwidth]{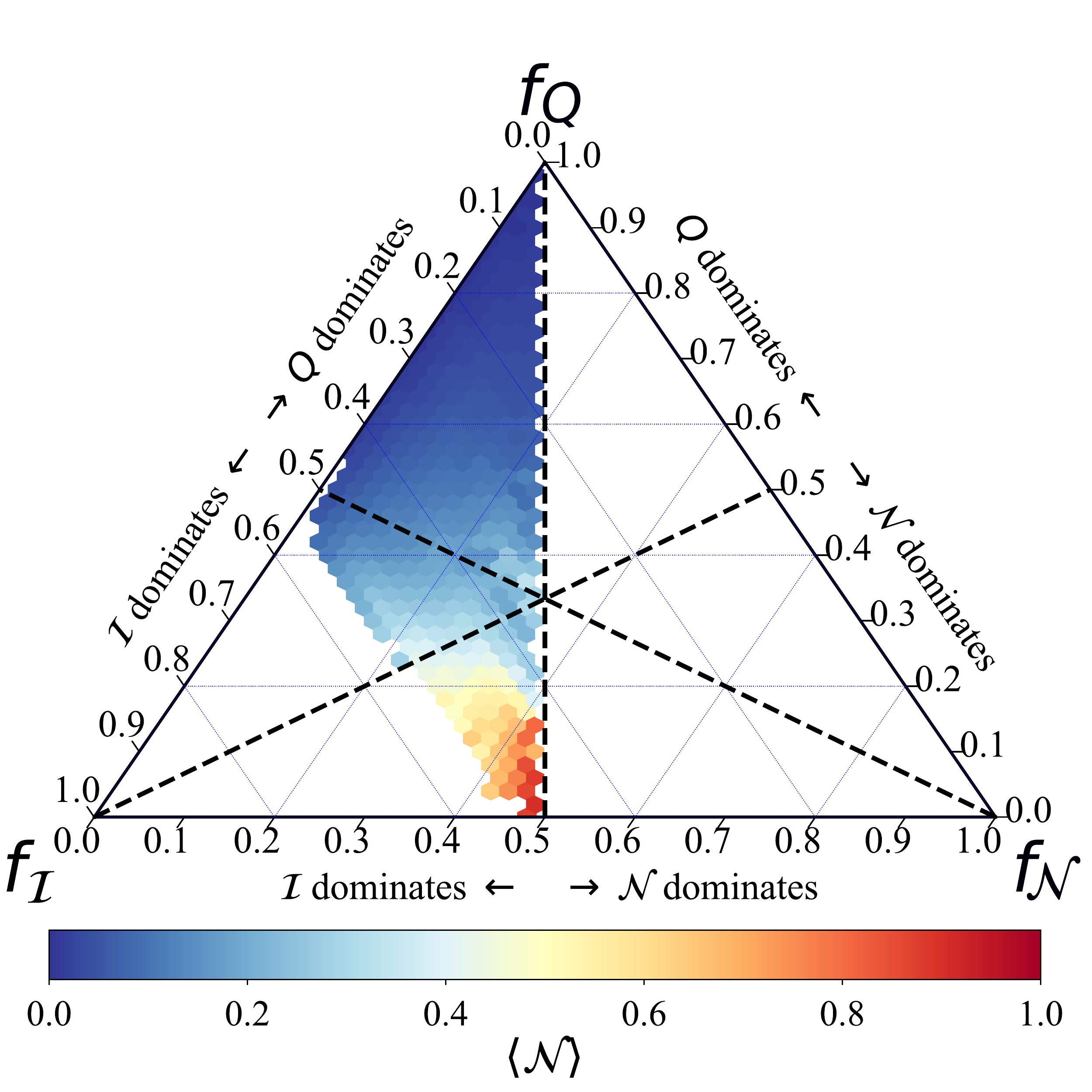}}{0.2in}{.05in}
\topinset{\bfseries(c)}{\includegraphics[width=0.5\columnwidth]{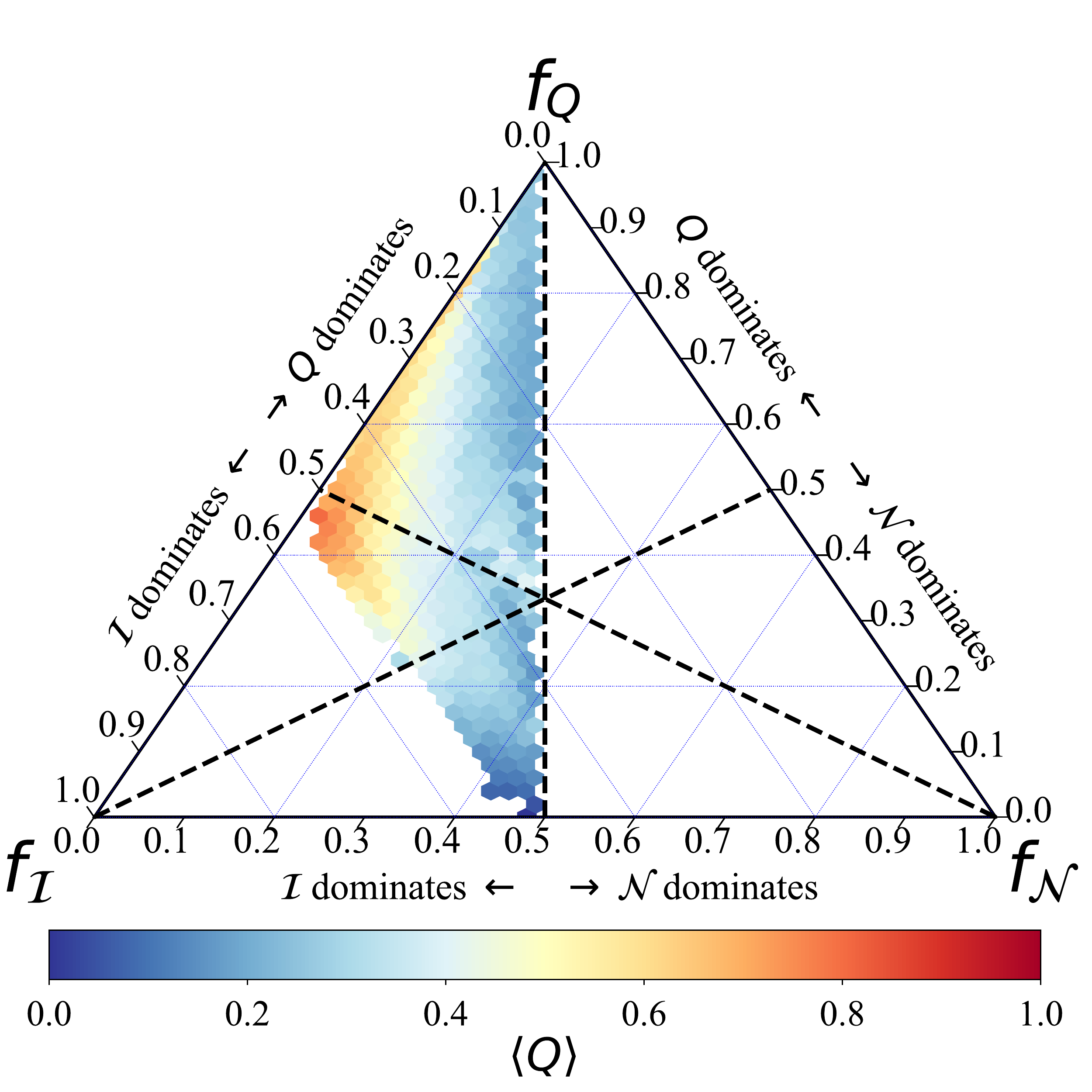}}{0.2in}{.05in}
\topinset{\bfseries(d)}{\includegraphics[width=0.5\columnwidth]{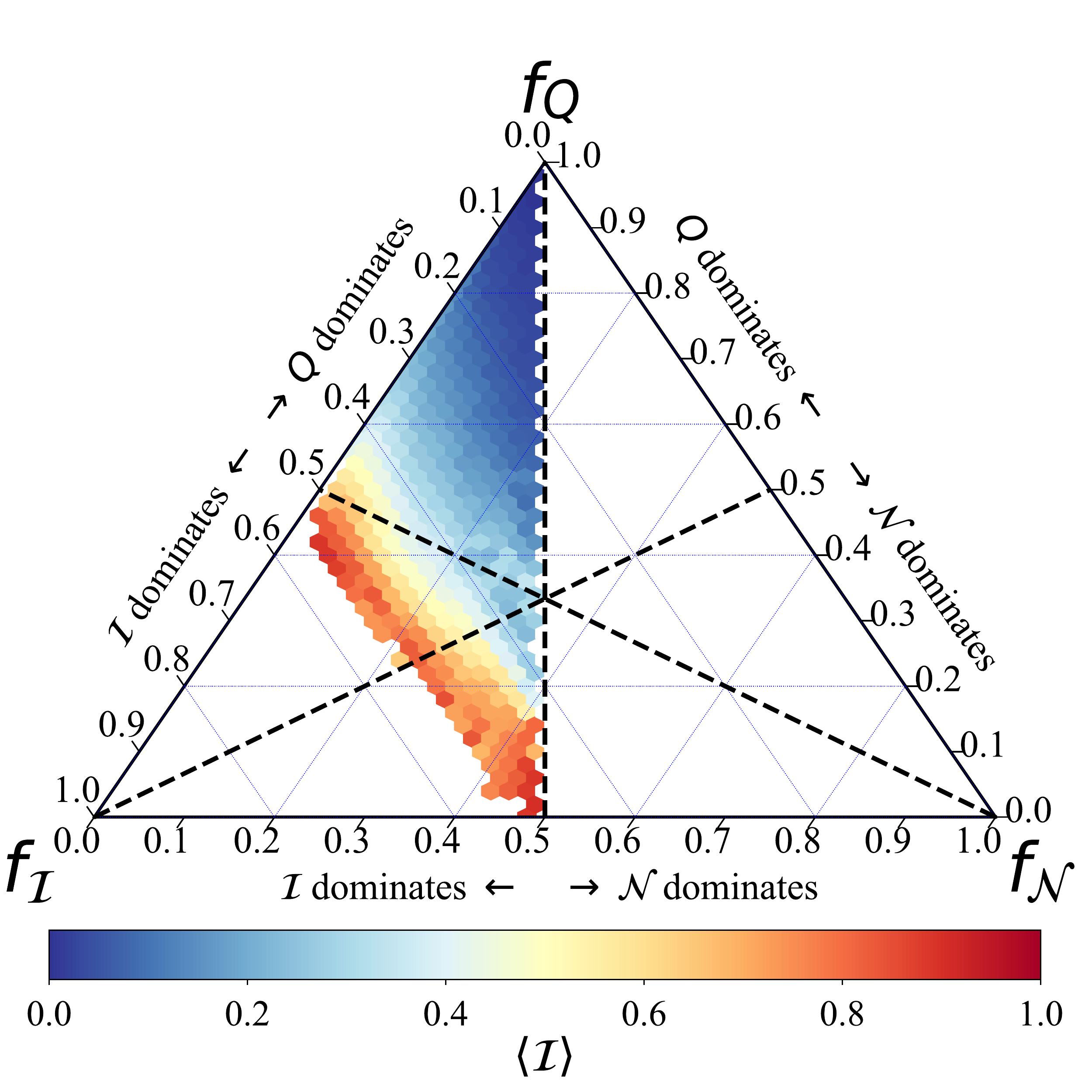}}{0.2in}{.05in}
\end{tabularx}
\caption{Panel (a) shows the distribution of the generated networks over the ternary plot. The color bar indicates the amounts of networks in each bin. Panels (b), (c) and (d) show the average absolute value of $\mathcal{N}$, $Q$ and $\mathcal{I}$, respectively.} \label{fig:heatmaps_measures}
\end{figure*}

We have measured $\mathcal{N}$, and optimized $Q$ and $\mathcal{I}$ for more than $2\times10^5$ networks on a wide range of parameters: $B\in [1,9]$; $\xi \in [1.5,7]$; $p\in [0,0.6]$; and $\mu\in[0,0.6]$. Restricting $p$ and $\mu$ to $0.6$ allows us to introduce a high level of noise while still preserving some underlying mesoscale structure. For modularity and in-block nestedness optimization, we have used the extremal optimization algorithm \cite{duch2005community}, adapted to the corresponding objective functions (Eqs.~\ref{eq:mod} and \ref{eq_inb}, respectively). Community size of $N_B= 50$ was assumed, so as we add communities we are also increasing the size of the network, the total number of nodes being $N_{T}=BN_B$. Fixing $N_{T}$ and reducing the size of the communities while increasing $B$ produces equivalent results. 

Figure~\ref{fig:heatmaps_measures} shows the results represented over four ternary heatmap plots. A ternary plot, or simplex, is a three-variable diagram in which the sum of the variables is a constant --1 in this case-- (for further details cf.~Supplemental Material). Panel (a) shows a density plot of the generated networks over the simplex, and the colorbar indicates the amount of networks in each bin of the ternary plot. Relying on these results, it is apparent that the most frequent architecture is predominantly modular. This is expected since most generated networks have blocks ($B>1$), and in-block nested networks are more restrictive, in terms of internal organization, than modular ones. The color code in panels (b), (c) and (d) reports the average absolute value of $\mathcal{N}$, $Q$ and $\mathcal{I}$. Dashed black lines have been added as visual aids to evince \textit{dominance} regions.
A quick glance already shows that the highest values of $\mathcal{N}$ and $Q$ never overlap, while $\mathcal{I}$ bridges between them. This is a valuable insight for the analytical results in the remainder of the article.

Another outstanding feature in Fig.~\ref{fig:heatmaps_measures} is the existence of sharp boundaries in the ternary plots. The first boundary, $F_{1}$ in panel (a), is induced by the definitions of $\mathcal{N}$ and $\mathcal{I}$: as stated, Eq.~\ref{eq_inb} reduces to Eq.~\ref{eq_nest} when $B=1$. Translated to coordinates on the simplex, $F_{1}$ simply reflects that the contribution of $\mathcal{N}$ is always equal or smaller than the contribution of $\mathcal{I}$, $f_{\mathcal{N}} \le f_{\mathcal{I}}$. More interesting, however, is the existence of $F_{2}$, which suggests that there is an inherent limit that prohibits in-block nestedness to dominate further over $Q$. On close inspection (see Fig.~S3 of the Supplemental Materials), networks which map onto $F_2$ have high values of $\xi$ and very low values of $p$ and $\mu$. We build on this finding to construct our analytical approach below.


The specific configuration of parameters along $F_{2}$ in Fig.~\ref{fig:heatmaps_measures} points at a well-defined family of network configurations: a ring of star graphs, $G_{s}$ hereafter. Indeed, an extreme shape parameter ($\xi \to \infty$), perfectly nested intra-block structure ($p=0$) and minimum inter-block connectivity ($\mu \approx 0$) to guarantee a single giant component, render a network model which depends only on $B$, ranging from a single star ($B=1$) to a set of stars ($B>1$) connected with a single link through their central nodes. In other words, $G_s$ provides the closest compatible network architecture for the boundary $F_2$. 
Given a ring of star graphs with $B$ communities and $N_B$ nodes per community, we can analytically derive the exact values $\mathcal{N}$, $\mathcal{I}$ and $Q$.

\paragraph*{Nestedness.} We obtain the analytical expression for $\mathcal{N}$ from the expression in Eq.~\ref{eq_nest}. The pair overlap of a generalist node (the center of each star subgraph), $g$, with a specialist node (periphery of a star), $s$, is $\mathit{O}_{gs}/k_s = 1$ if $g$ and $s$  belong to the same star (and 0 otherwise). For all those pairs (regardless of the star they belong to), the null model contribution is $\langle\mathit{O}_{gs}/k_s\rangle = (N_B+1)/BN_B$. We can obtain in a similar way the terms for the the generalist-generalist pairs between stars. Summing up all the contributions, the final expression for $\mathcal{N}$ is:
\begin{equation}
\mathcal{N}=\frac{B N_B^{3} - BN_{B}^2  - 3BN_B + B + 2N_B+2}{B N_B\left(BN_B^2 + BN_B -N_B -1\right)}.
\label{eq_nest_stars}
\end{equation}

\paragraph*{Modularity.} While the optimal partition for an arbitrary network cannot be easily obtained, this is not the case for $G_s$ where each star in the ring forms a community.
Thus, we can easily derive the contribution of each star to the total $Q$ following Eq.~\ref{eq:mod}. The first element is $l_c = N_B - 1$. 
The second element (the amount of links of the network) includes links within and between communities, $L = B(N_B-1) + B= BN_B$. The last term, the sum of the degrees of all the nodes in community $c$, corresponds to $d_c = 2N_B$. Assembling these, we obtain the modularity of $G_s$ as
\begin{equation}
\begin{aligned}
 Q&=B\left[ \frac{N_B -1}{BN_B}- \left(\frac{2N_B}{2BN_B}\right)^2 \right] &= 1- \frac{1}{N_B} - \frac{1}{B},
 \end{aligned}
 \label{eq_mod_stars}
\end{equation}
which is equivalent to the general expression derived in \cite{fortunato2007resolution}. 

\paragraph*{In-block nestedness.} The derivation of $\mathcal{I}$ resembles that of $\mathcal{N}$, with the difference that only nodes within the same community contribute; thus, all stars have the same contribution. Focusing now on each star, we have only two contributing terms to the sum: the pair overlap between specialist nodes, $s$, and the pair overlap of the generalist node, $g$, with the specialists. In both cases, the contribution is 1. The null model corrections are $\langle \mathit{O}_{gs}\rangle={k_gk_s}/{BN_B}=(N_B+1)/BN_B$ and $\langle \mathit{O}_{ss}\rangle={k_sk_s}/{BN_B}={1}/{BN_B}$. Finally, the size of the communities is $C_g=C_s=N_B$. Replacing all the contributions in Eq.~\ref{eq_inb}, we obtain 
\begin{equation}
\begin{aligned}
\mathcal{I} &= 1-\frac{3}{BN_B} -\frac{2}{N_B}.
\end{aligned}
\label{eq_ibn_stars}
\end{equation}

All the expressions presented above were obtained considering a closed ring, on which the number of intercommunity links is $B$. For the cases $B=1$ and $B=2$, the number of intercommunity links is $B-1$ and the degree of the generalist nodes is $k_g=N_B-1$ and $k_g=N_B$, respectively (cf.~Supplemental Materials for details on these cases).

We now focus on the bounds that $\mathcal{N}$ and $Q$ impose on each other in some important limits. These correspond to scenarios in which the number of blocks, $B$, and the size of the blocks, $N_B$, tend to $\infty$. 

We start with $N_B \to \infty$. In this case, Eqs.~\ref{eq_nest_stars} and \ref{eq_mod_stars} reduce to
\begin{align}
 	\displaystyle\lim_{N_B \to \infty} \mathcal{N} =  \frac{1}{B}, &&   \displaystyle\lim_{N_B \to \infty} Q = 1 - \frac{1}{B},
 \label{eq_nb}
\end{align}
which implies that, under these circumstances, $\mathcal{N}$ and $Q$ are complementary --in accordance with the empirical results in \cite{borge2017emergence}. This result proves analytically the antagonism that exists between these two structural patterns.

With respect to the case  $B \to \infty$, Eqs.~\ref{eq_nest_stars} and \ref{eq_mod_stars} turn now
\begin{align}
 	\displaystyle\lim_{B \to \infty} \mathcal{N} =  0, && \displaystyle\lim_{B \to \infty} Q = \frac{N_B-1}{N_B}.
 \label{eq_b}
\end{align}
These results fit the expectation that, with increasing $B$, the negative contribution of non-overlapping nodes through the null model overcomes the decreasing positive contributions realized.

Finally, with respect to in-block nestedness, the analytical calculations in both limits yield
\begin{align}
\displaystyle \lim_{N_B \to \infty} \mathcal{I} = 1 && \displaystyle\lim_{B \to \infty} \mathcal{I}  =  \frac{N_B-2}{N_B}
 \label{eq_both}
\end{align}

\begin{figure}[h!]
\centering
    	\begin{tabular}{l}
                \includegraphics[width=\columnwidth]{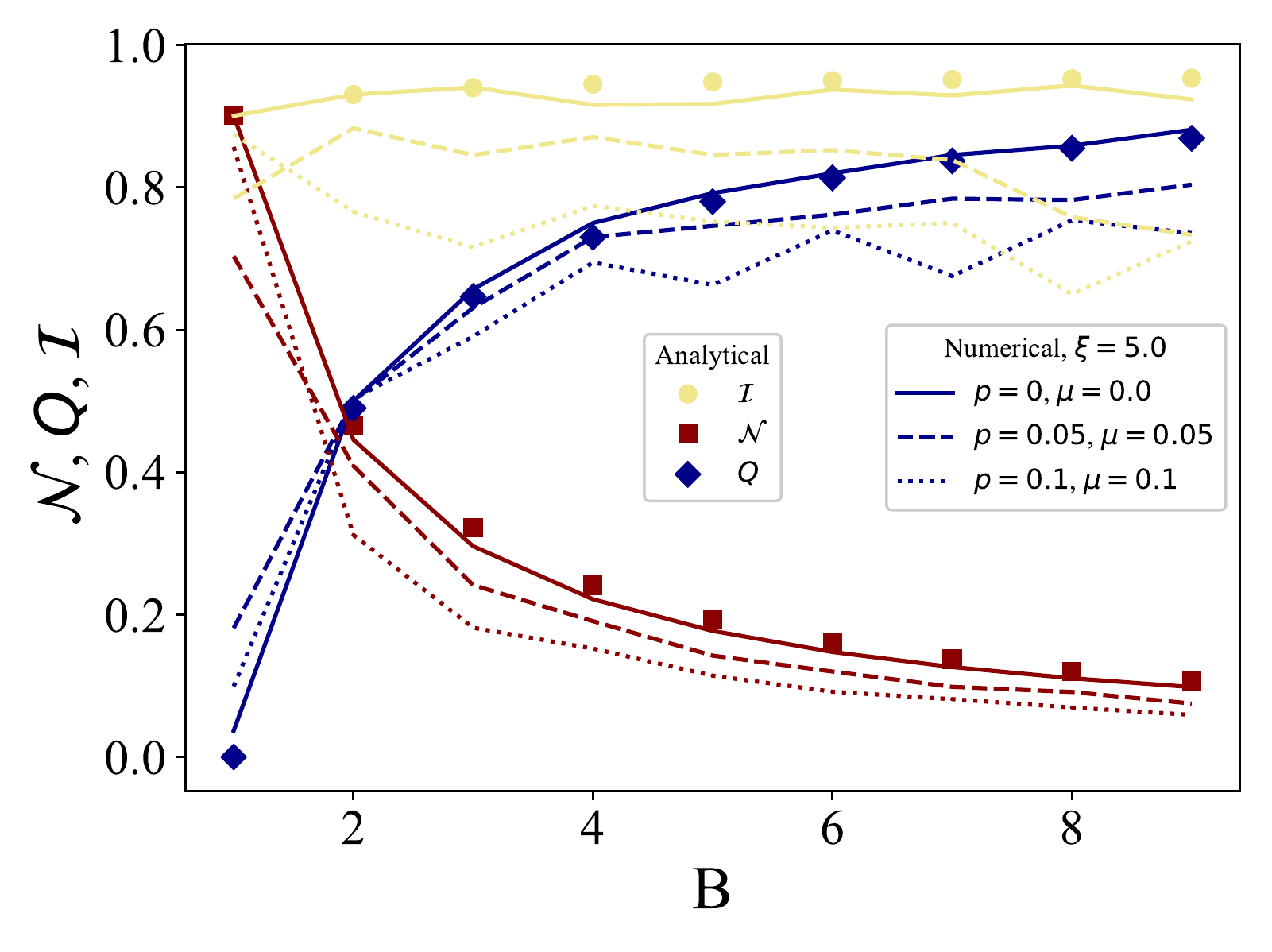}                 
        \end{tabular}
        \caption{Comparison of the analytical (symbols) and numerical (lines) values of $\mathcal{N}$, $Q$, $\mathcal{I}$  with respect to $B$. All the calculations were performed by taking  $N_B=50$ and $\xi=5$. The values for $p$ and $\mu$ parameters are indicated in the plot legend.}
        \label{fig:N_Q_I_vs_B}
\end{figure}

Noteworthy, the generated synthetic networks follow closely the predictions in limiting cases: Fig.~\ref{fig:N_Q_I_vs_B} reports the analytical estimation of $\mathcal{N}$, $Q$ and $\mathcal{I}$ (Eqs.~\ref{eq_nest_stars}-\ref{eq_ibn_stars}, symbols in the figure) against $B$, and the numerical results for networks generated under different parameters. As the generated networks deviate from the ring of stars $G_s$ (i.e.~$p>0$ and $\mu>0$), results show a worse fit to the analytical prediction. The mutual bounds that $\mathcal{N}$ and $Q$ impose on each other are obvious, observing a perfectly anti-correlated behavior between nestedness and modularity. Finally, as the networks transition from a nested ($B = 1$) to a modular ($B > 1$) architecture, the values of in-block nestedness remain very high and almost constant.


\begin{figure*}
\centering
\def\stackalignment{l}
\topinset{\bfseries(a)}{\includegraphics[width=0.5\textwidth]{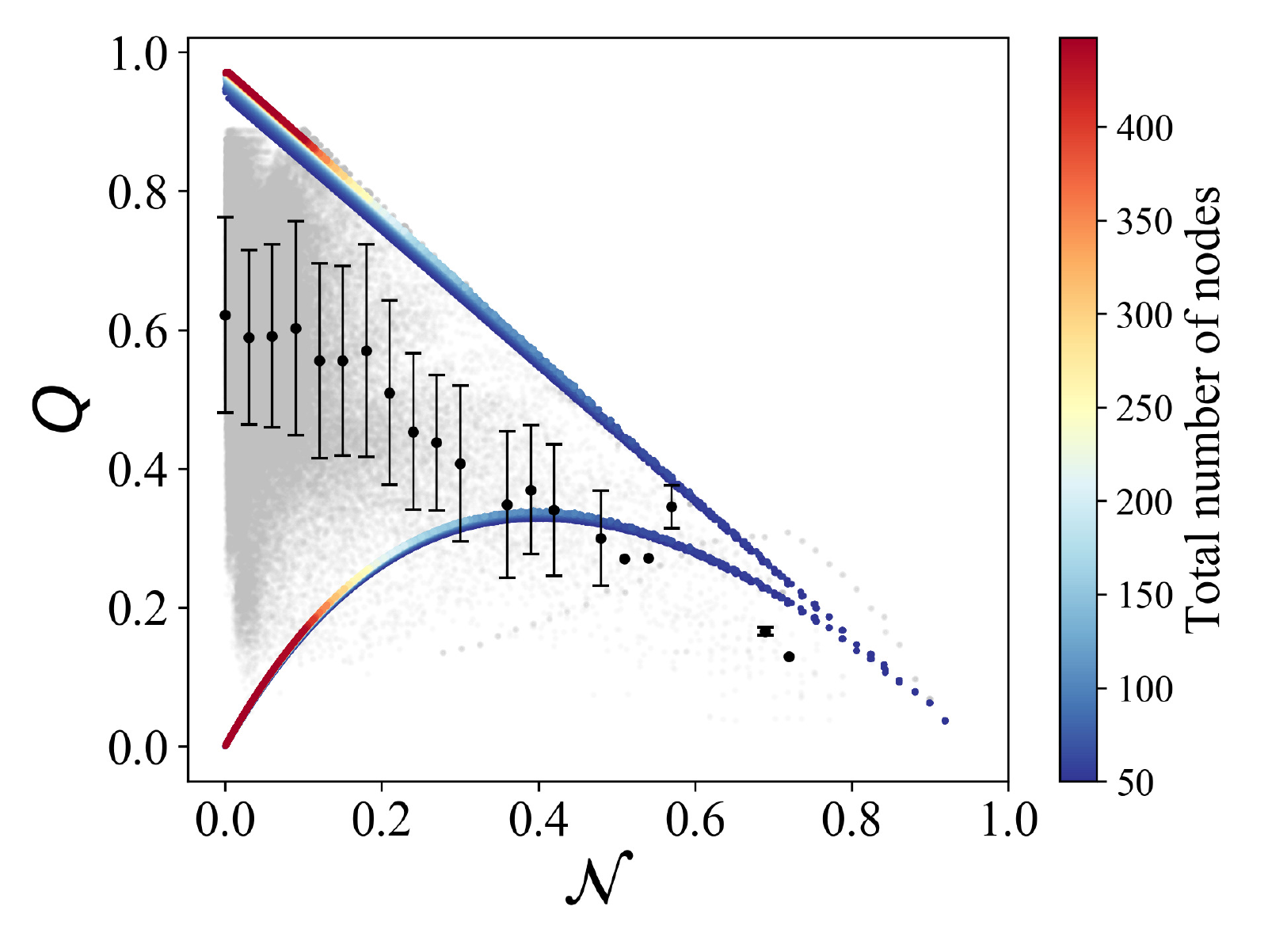}}{-0.1in}{0.05in}
\topinset{\bfseries(b)}{\includegraphics[width=0.43\textwidth , height=0.285\textheight]{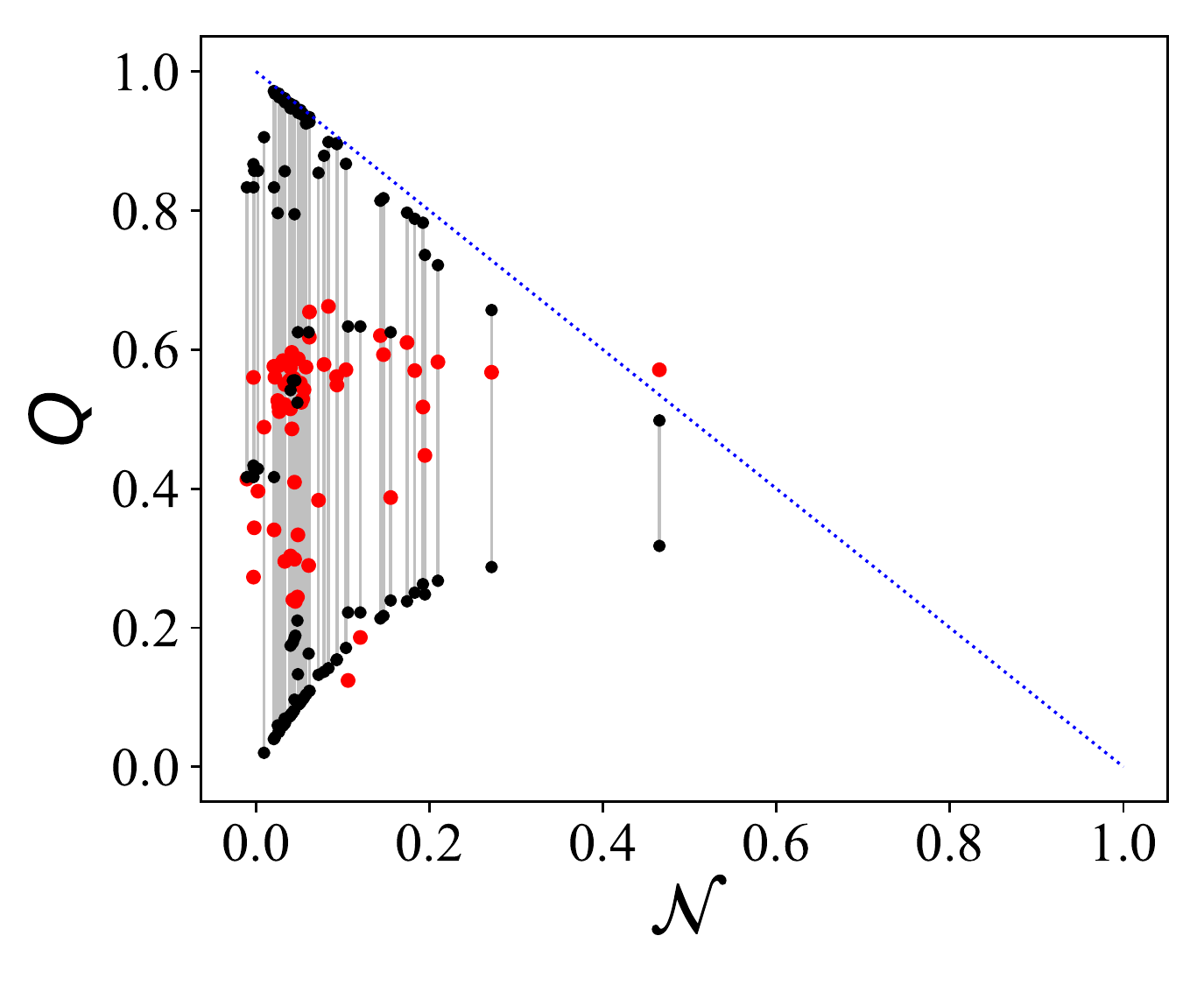} }{-0.1in}{0.05in}
\caption{Panel (a) show the values of $Q$ obtained after optimization (grey dots), plotted against $\mathcal{N}$ for over $2\times10^5$ generated networks and  panel (b) for the set of unipartite social network (57 networks) analyzed in \cite{sole2018revealing}. Colored upper and lower bounds of $Q$ have been obtained from Eqs.~\ref{eq_mod_stars}-\ref{eq_ibn_stars} and $B^{*} = f(\mathcal{N},N_{T})$. The color bar indicates the network's size.}\label{fig:Q_vs_N}
\end{figure*} 

The previous results open a new front to understand the co-occurrence of macro- and mesoscale patterns in complex networks. Complementary to the inherent limits of $Q$ \cite{fortunato2007resolution,lancichinetti2011limits}, we have now evidence that a certain connectivity arrangement (i.e.~nestedness) places hard limits to modularity, at least in extreme settings. This certainty paves the way to obtain estimations for $Q$ prior to computationally costly endeavors: indeed, relaxing those conditions to realistic parameters, soft bounds for $Q$ may be defined. The derivation of these bounds are presented below.

We start from a $G_s$ of size $N_{T} = BN_B$. From here, Eq.~\ref{eq_nest_stars} can be rewritten in terms of $N_{T}$ and $B$, and an estimation on the number of blocks as a function of $\mathcal{N}$ and $N_{T}$ can be obtained, i.e.~$B^{*}=f(\mathcal{N},N_{T})$. The upper bounds for $Q$ and $\mathcal{I}$ are thus readily available, applying $B^{*}$ to Eqs.~\ref{eq_nest_stars}-\ref{eq_mod_stars}, i.e.~assuming that the network structure lies on the boundary $F_{2}$. With the actual measure of $\mathcal{N}$, and upper estimations for $Q$ and $\mathcal{I}$, we can obtain the relative fraction that each measure contributes to the ternary plot along the $F_2$ boundary, $f_\mathcal{N}^{\uparrow}$, $f_Q^{\uparrow}$ and $f_\mathcal{I}^{\uparrow}$. 

To obtain the lower bounds for $Q$, we observe from  Fig.~\ref{fig:heatmaps_measures}(b) that, if we move from boundary $F_2$ to boundary $F_1$ on the $f_{Q}$-axis direction, i.e.~horizontally in Fig.~\ref{fig:heatmaps_measures}, the $\mathcal{N}$ values are approximately constant with respect to the contributions of $Q$ ($f_Q^{\uparrow}$ and $f_Q^{\downarrow}$). This allows us to make an approximation for the contributions of $Q$ in the ternary plot at $F_1$ as $f_Q^{\downarrow} \approx f_Q^{\uparrow}$. Additionally, we know that $\mathcal{N}=\mathcal{I}$ at boundary $F_1$. Thus, 
\begin{eqnarray}
f_Q^\downarrow = \frac{Q}{Q+\mathcal{I}+\mathcal{N}} = \frac{Q}{Q+2\mathcal{N}},
 \label{eq_cq}
\end{eqnarray}
from which a lower bound for $Q$ can be obtained. 

Figure \ref{fig:Q_vs_N} shows the values of $Q$ as a function of  $\mathcal{N}$ for the previous synthetic ensemble ($\sim2\times10^5$ networks; panel (a), grey dots); and 57 real unipartite  networks (panel (b), red dots)  \cite{sole2018revealing}. In panel (a), the values of the theoretical upper and lower bounds are plotted in colors, the color bar indicating the network size. Our approximation of $Q$ bounds is in good agreement with actual values obtained after optimization: most of the optimized $Q$ values lie within the estimated soft bounds. Despite the wide range of parameters --clearly far from limiting cases-- estimated upper bounds behave like $Q = 1- \mathcal{N}$ almost perfectly. While these bounds are trivial for $\mathcal{N} \approx 0$, we observe that intermediate values of nestedness provide relevant information about the possible mesoscale organization of the network. $Q$ values above the upper bound correspond to networks with a single community $B=1$ and perfectly nested structure, $p=0$ (see Fig.~S4). These networks --less than 0.1\% of the total-- are dense enough to allow a partition with $B>1$ where the nodes of higher degree are gathered in a block, resulting in values of $Q$ larger than expected \cite{sole2018revealing}. Values below the lower bound approximation are more numerous --although still a small fraction of the total. This imprecision shows that there is room to improve the underlying assumption, i.e.~that $\mathcal{N}$ values are constant with respect to the contributions of $Q$. In the same spirit, upper and lower bounds for $\mathcal{I}$ can be as well approximated from the actual value of $\mathcal{N}$ (see Fig.~S5). For the sake of completeness, $Q$-$\mathcal{I}$ scatter plots are shown in Fig.~S6, where we see that $\mathcal{I}$ and $Q$ can coexist, i.e.~there is no clear map from one to the other.
Remarkably, bound estimation for real networks in Fig.~\ref{fig:Q_vs_N}(b) closely follows the approximation for synthetic networks: the inferior and superior trends of black dots are the predicted  lower and upper bounds. It is worth highlighting, that bound estimation --which has a very low computational cost-- renders non-trivial information for some networks ($\mathcal{N} \gtrsim 0.2$).


While the study of macro- and mesoscale arrangements in complex networks has been studied in depth, we know little about how they affect each other. Understanding and, above all, quantifying such pattern interactions becomes necessary for many reasons. First, because empirical evidence suggests the concurrence of more than one pattern within the same network \cite{olesen2007modularity,fortuna2010nestedness,flores2011statistical,borge2017emergence}. Second, because a preliminary approximation of the mesoscale structural features of a network is appealing, at the face of prohibitive costs to analyze very large amounts of data. Further, the interplay between nestedness and modularity is thought fundamental to decipher the dynamical behavior of many empirical systems (like ecological, economic, and technological networks among others). 
In this work, we have quantified, numerically and analytically, the interference between nestedness (at the macroscale) and modularity and in-block nestedness (at the mesoscale). We show that modularity and nestedness are antagonistic architectures: the growth of one implies the decline of the other, and bounds to modularity can be estimated even in far from idealized settings. Intermediate nested-modular regimes are possible, pointing directly at in-block nested structures as the natural transition between the other two.


\bibliography{literature}

\begin{thebibliography}{41}
\bibitem [{cit()}]{cite-foot}%
  \BibitemOpen
  \href@noop {} {}\bibinfo {note} {Without loss of generality, in the following
  we consider purely unipartite networks. The extension to bipartite ones is
  trivial.}\BibitemShut {Stop}%
\end{thebibliography}

\end{document}


\title{Supplemental Material for\\ ``Antagonistic Structural Patterns in Complex Networks''}

\author{{ Mar\'ia J. Palazzi$^1$, Javier Borge-Holthoefer$^1$, Claudio J. Tessone$^2$ and Albert Solé-Ribalta$^1$}\\
{\small \em $^1 $Internet Interdisciplinary Institute (IN3), Universitat Oberta de Catalunya, Barcelona, Catalonia, Spain\\
$^2$URPP Social Networks, Universit\"at Z\"urich, Switzerland.}}

\date{\today}
\maketitle

The following section provides a detailed description of the  network generation model employed to carry out the experimentally explore the structural relationship between nestedness, modularity and in-block nestedness. The model is a version of the one developed by  Sol\'e-Ribalta {\it et al.} in \cite{sole2018revealing}  modified to generate size increasing networks with a fixed block size, instead of networks with fixed size, as in the original formulation.

\section{Supplemental Section I: Probabilistic model for the synthetic network generation} \label{toy_model_extended}

We generated synthetic in-block nested networks employing the benchmark graph model introduced in \cite{sole2018revealing}. The model is implemented in terms of links probabilities, and asks for four parameters: the number of modules $B$, the fraction of inter modules links $\mu$, the fraction of links outside the perfect nested structure $p$, and the shape parameter $\xi$, that represents the slimness of the nested structure. 

The perfect nested structure is generated using a function inspired in the $p$-norm ball equation, written as
\begin{equation}
f_n(x)=1-(1-x^{1/\xi})^{\xi},
\label{unit}
\end{equation}
where $ \xi \in [1, \infty)$ and $ x \in [0, 1]$. The perfect nested structure is constructed by adding a link into each matrix position whose center lies above the curve in Eq. \ref{unit}, resembling an upper left triangle. Afterwards, starting with a fixed community size and given a number of communities $B$, we construct the adjacency matrix of a network by building $\lfloor B\rfloor$  blocks of size $n_r$  and a remaining block of size $\{B\} n_r$, such that, the total size of the network is $N_r= Bn_r $, forming a block diagonal matrix\footnote{$\lfloor \cdot \rfloor$ refers to the integer part function}$^{,}$\footnote{$\{ \cdot \}$ refers to the fractional part function}.

The probability of having a link between two nodes $i$ and $j$ inside a block is given by 
\begin{equation}
P(A^c_{ij})=[(1-p+pp_{r})\Theta(jN_r-f_n(iN_r))+ p_r (1-\Theta(jN_r -f_n(iN_r))](1-p_i),
\end{equation}
The term within square brackets is related to the intra-block noise,  being $p$ the probability of removing a link from the perfect nested structure. Hence, the term $(1 - p)$ corresponds to the probability of not altering the link. The second, $pp_{r}$, corresponds to the probability of recovering a link after removal and $p_r = pE(N_r - E + pE)^-1$, corresponds to the probability of selecting link $A_{ij}$  from the distribution of removed links, where $E$ is the number of links in the network. The first two terms are restricted by the Heaviside function $\Theta$, to the region of perfect in-block nestedness. Finally, the term $(1 - p_i)$ corresponds to the probability of not removing the link in the process of generating inter-block noise, where $p_i={\mu(B-1)}/{B}$ and  $ \mu \in [0, 1]$.

Finally, the probability of a inter-block link is given by,
\begin{equation}
P(A^o_{ij})=  \frac{2Ep_i}{2(B-1)N_r^2} =\frac{\mu E}{N_r^2B},
\end{equation}
the numerator corresponds to the number of removed links from the blocks, and the denominator corresponds to the possible places where each of those links can be relocated. 

\begin{figure}[h]
\includegraphics[width=0.5\columnwidth]{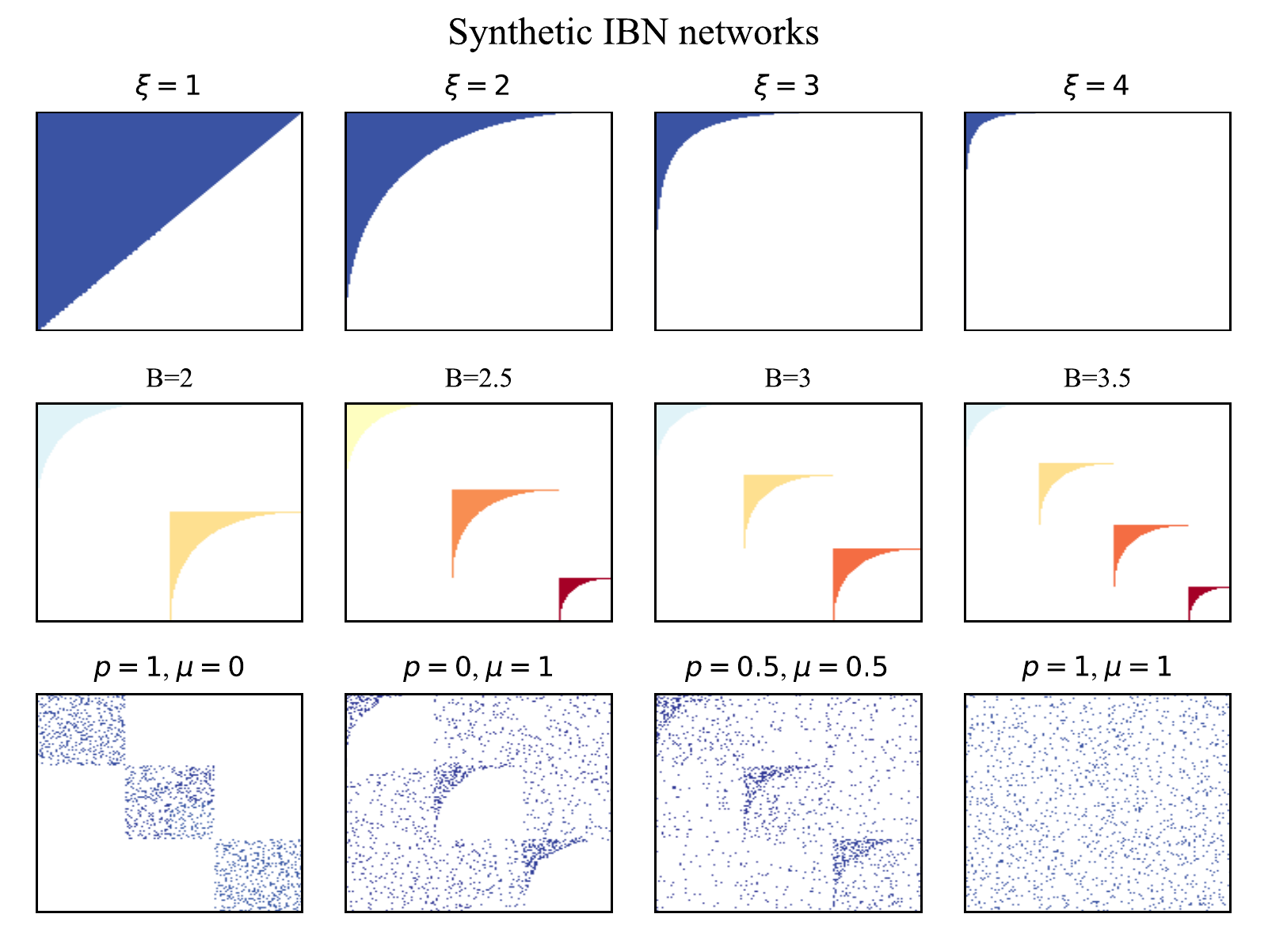}
\caption{Examples of synthetic network generation with the model introduced in \cite{sole2018revealing}. The top and middle rows show the effects of the shape parameter $\xi$ and the number of blocks $B$, respectively, in a noiseless scenario ($p=\mu=0$). The bottom row provides some examples of the effect of the noise parameters $p$ and $\mu$.}
\label{fig_toy}
\end{figure}

Figure~\ref{fig_toy} shows several examples of the networks the model is able to generate. Perfectly nested networks are generated with $B=1$ and varying values of $\xi$, then fixing $p=\mu=0$. With the same settings and $B > 1$, perfect in-block nested networks are obtained. Ideal modular networks are generated fixing $\mu=0$ and varying values of $p$ and $\xi$. Parameters $p=\mu=1$ generate Erd{\H{o}}s-R\'{e}nyi networks, regardless of $B$ (bottom-right).

\newpage

\section{Construction and reading of the ternary plot}
A ternary plot or simplex is a three-variable diagram on which each point in the plot represents the proportions between the three considered variables, and is obtained as $f_{x}=\frac{x}{x+y+z}$, $f_{y}=\frac{y}{x+y+z}$ and $f_{z}=\frac{z}{x+y+z}$. In our case, each axis corresponds to the fractional values of the structural patterns we analyze: the bottom-left vertex represents purely in-block nested networks (fractional values: $f_{\mathcal{N}}=f_{Q}=0$, $f_{\mathcal{I}}=1$), the bottom-right represents networks that are purely nested ($f_{\mathcal{N}}=1$, $f_{Q}=f_{\mathcal{I}}=0$), and the top vertex those networks that are purely modular ($f_{\mathcal{N}}=f_{\mathcal{I}}=0$, $f_{Q}=1$). This representation is convenient to explore which network structural configurations map onto which region on the ternary plot. 

\begin{figure}[h]
\includegraphics[width=0.4\columnwidth]{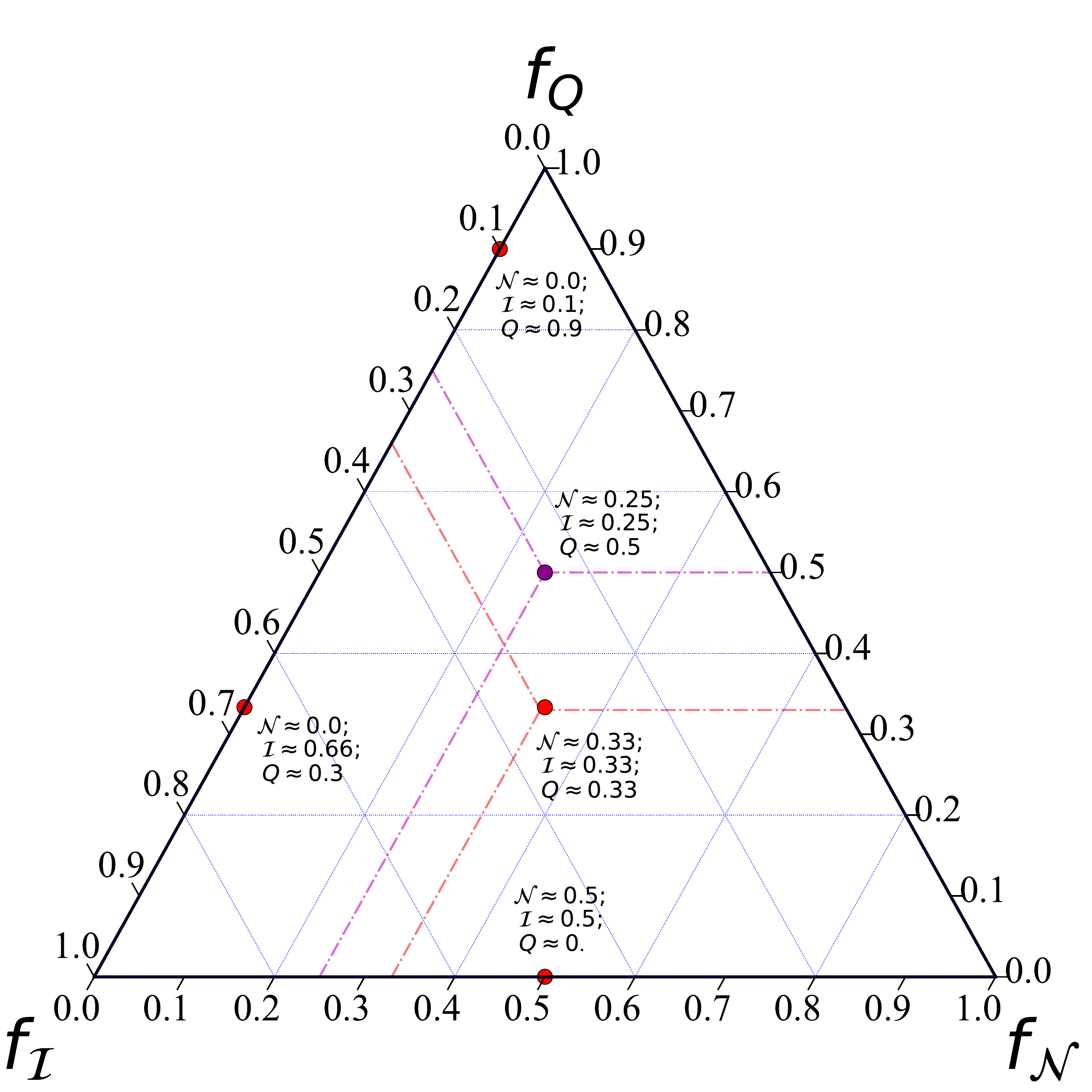}
\caption{Representation of the three variables in the ternary plot showing exemplar points with different proportions.}
\label{fig_ternary_example}
\end{figure}

\section{Supplemental Section II: Analytic expressions along $F_2$ for the cases of $B=1$ and $B=2$}
 \label{analytic_special_cases}
This section provides a complementary formulation for two particular cases of the ring of star graphs, $G_s$. In the main text, Eqs.~4-6 were developed for the general case $B > 2$. Here, we introduce as well the exact derivation of the expressions  $\mathcal{N}$, $Q$ and $\mathcal{I}$ for the cases of a single star ($B=1$) and two stars connected through their central nodes ($B=2$). As stated in the main text, for these two situations, we have to take into account the change in the number of inter-community links and the degree of the generalist nodes.

\subsection{Nestedness}

\subsubsection{A single star graph, $B=1$}
The computation of the pair overlap for the evaluation of nestedness when $B=1$ requires only the following terms: the pair overlap of a generalist node (the center of each star subgraph), $g$, with the specialist nodes $s$ which is $\mathit{O}_{gs}/k_s = 0$; and the pair overlap between all the specialists nodes $\mathit{O}_{ss}/k_s = 1$, the degree of the generalist node is $k_g=N_B-1$ and the null model corrections $\langle \mathit{O}_{gs}\rangle={k_gk_s}/{BN_B}=(N_B-1)/BN_B$ and $\langle \mathit{O}_{ss}\rangle={k_sk_s}/{BN_B}={1}/{BN_B}$
 
\begin{equation}
\begin{aligned}
\mathcal{N} &= \frac{2}{N_{B}(BN_{B} - 1)} \left \{  \left[ - \frac{N_B-1}{BN_B}(N_B-1) \right]+  \left[  \left(1 - \frac{1}{BN_B} \right)\frac{(N_B-2)(N_B-1)}{2}\right] \right \}\\ &=\frac{(N_B-2)(N_B-1)}{N_B^2} - \frac{2(N_B-1)^2}{N_B^2(N_B-1)},
\end{aligned}
\end{equation}%

\subsubsection{Two star graph, $B=2$}
For this scenario we have to take into account the change on the degree of the generalist node $k_g=N_B$ and additional terms such as: the pair overlap between the two generalists $\mathit{O}_{gg}/k_g$, the pair overlap between a generalist with the specialist from the other community $\mathit{O}_{gs_{out}}/k_s$, and the pair overlap between a specialist with the specialists from the other community$\mathit{O}_{ss_{out}}/k_s$. Finally, we obtain 
\begin{equation}
\mathcal{N}= \frac{BN_B^3 - BN_B^2 - 6N_B^2 + 8N_B - 3}{BN_B^2(BN_B - 1)}
\label{eq_nest_stars}
\end{equation}
\subsection{Modularity}
\subsubsection{A single star graph, $B=1$}

Starting with the equation for modularity expressed as sum over the communities (Eq. 2 of the main text), we obtain the total number of links in the network and the number of links per community for a single star graph $N_B -1$, and the sum of the degrees of the nodes in the community $d_c=2(N_B -1)$. Now, we have the maximum modularity for $B=1$.
\begin{equation}
 Q=B\left[ \frac{(N_B -1)}{(N_B-1)}- \left(\frac{2(N_B-1)}{2(N_B-1)}\right)^2 \right]=0,
 \label{eq_mod2}
\end{equation}
\subsubsection{Two star graph,  $B=2$}
When $B=2$ the total number of links changes to $L=B(N_B -1)+1$ and the sum of the degrees of the nodes in the community is $d_c=2(N_B -1)+1$. So we obtain 

\begin{equation}
\begin{aligned}
 Q &=B\left[ \frac{(N_B -1)}{B(N_B-1)+1}- \left(\frac{2N_B-1}{2(B(N_B-1)+1)}\right)^2 \right]\\  \\&=2\left[ \frac{(N_B -1)}{2(N_B-1)+1}- \left(\frac{2N_B-1}{2(2(N_B-1)+1)}\right)^2 \right] \\ \\&=\left[ \frac{(2N_B -2)}{2N_B-1}- \frac{1}{2} \right],
 \label{eq_mod2}
 \end{aligned}
\end{equation}

\subsection{In-block nestedness}
\subsubsection{A single star graph, $B=1$}
Once again, we know that we will have only two contributing terms to our sum; the pair overlap between specialists $(s)$ nodes and the pair overlap of the generalist $(g)$ node with the specialists. Additionally, we know that for this case the degree of the generalist node is $k_G=N_B-1$ and the rest of the terms are:  the number of specialists nodes $N_{s}=(N_B -1)$, the null model corrections $\langle \mathit{O}_{g,s}\rangle={k_gk_g}/{BN_B}=(N_B-1)/BN_B$ and $\langle \mathit{O}_{s,s}\rangle={k_sk_s}/{BN_B}={1}/{BN_B}$, an the size of the communities is $C=N_B$. Finally the analytical expression for in-block nestedness when $B=1$ reads, 

\begin{equation}
\begin{aligned}
\mathcal{I} &=\frac{2}{N_B} \left \{ \left[ \frac{-(N_B-1)/BN_B}{(N_B-1)}\left(N_B -1\right) \right] + \left[ \frac{1- 1/BN_B}{(N_B-1)}\frac{\left(N_B -2\right) \left(N_B -1\right)}{2} \right] \right \} \\ \\&  =\frac{2}{N_B} \left \{ \left[-\frac{(N_B-1)}{N_B}  \right] + \left[ \frac{(N_B-1)(N_B-2)}{2N_B} \right] \right \},
\label{eq_ibn_stars}
\end{aligned}
\end{equation}
\subsubsection{Two star graph, $B=2$}
Now, for $B=2$ we have that the degree of the generalist node is $k_G=N_B$. Substituting this term we obtain a new expression for the in-block nestedness as

\begin{equation}
\mathcal{I}  =\frac{2}{N_B} \left \{ \left[ -\frac{1}{N_B}  \right] + \left[ \frac{(2N_B-1)(N_B-2)}{4N_B} \right] \right \},
\label{eq_ibn_stars}
\end{equation}

\section{Supplemental Figure 3}\label{toy_model_heatmaps}
\begin{figure}[H]
\centering
\def\tabularxcolumn#1{m{#1}}
\begin{tabularx}{\linewidth}{@{}cXX@{}}
\def\stackalignment{l}
\begin{tabular}{c}
\def\stackalignment{l}
\topinset{\bfseries(a)}{\includegraphics[width=0.25\columnwidth]{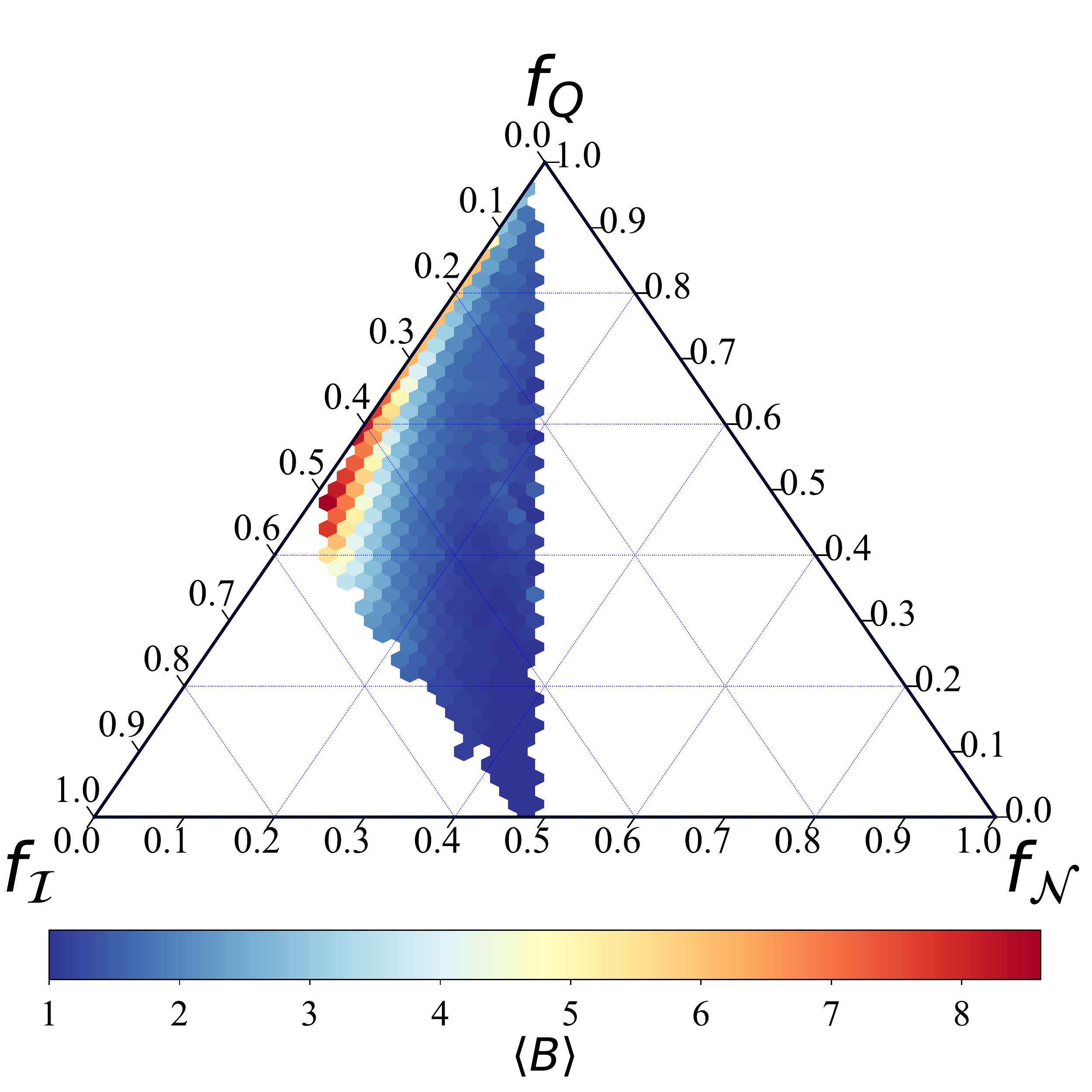}}{0.2in}{0.05in}
\topinset{\bfseries(b)}{\includegraphics[width=0.25\columnwidth]{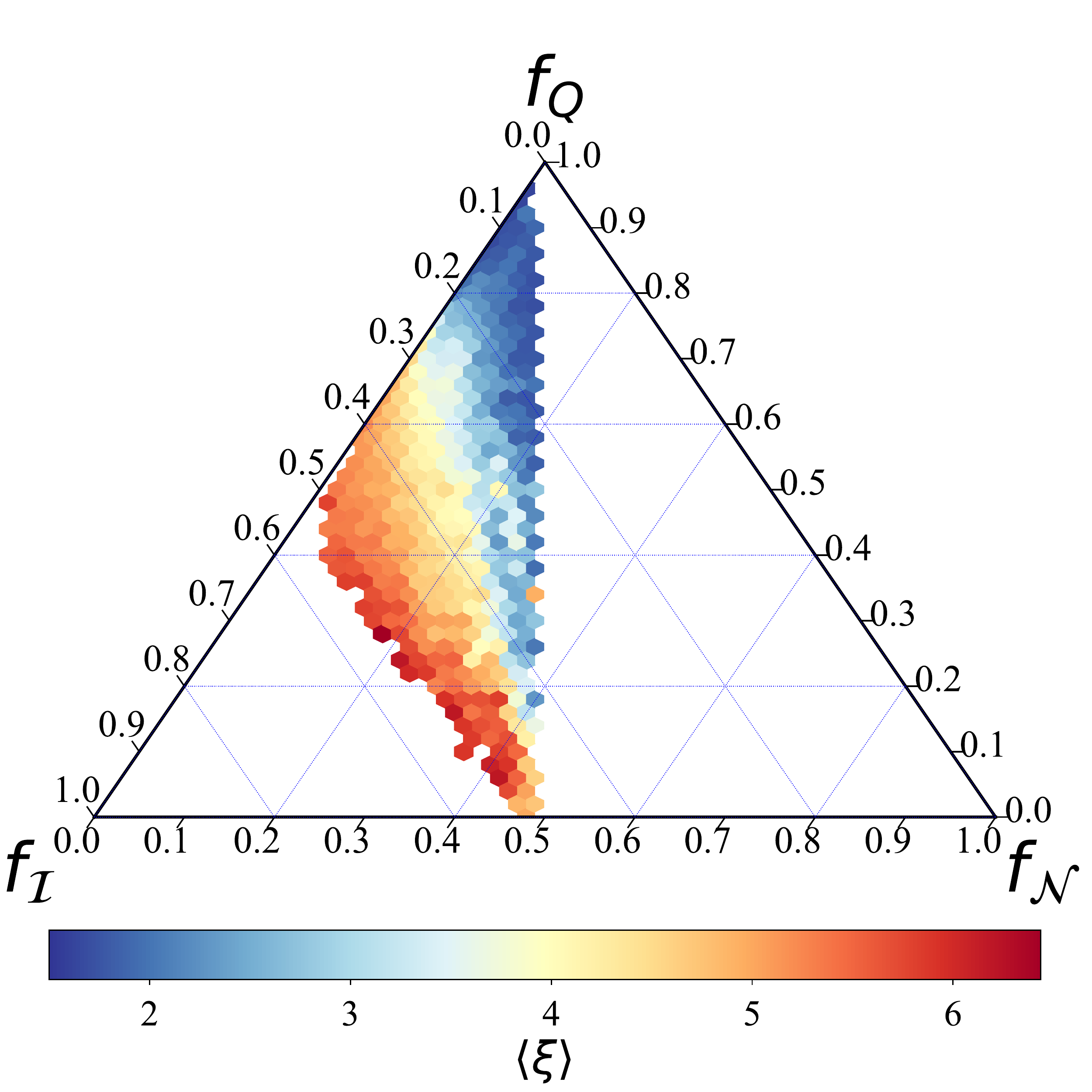}}{0.2in}{.05in}
\topinset{\bfseries(c)}{\includegraphics[width=0.25\columnwidth]{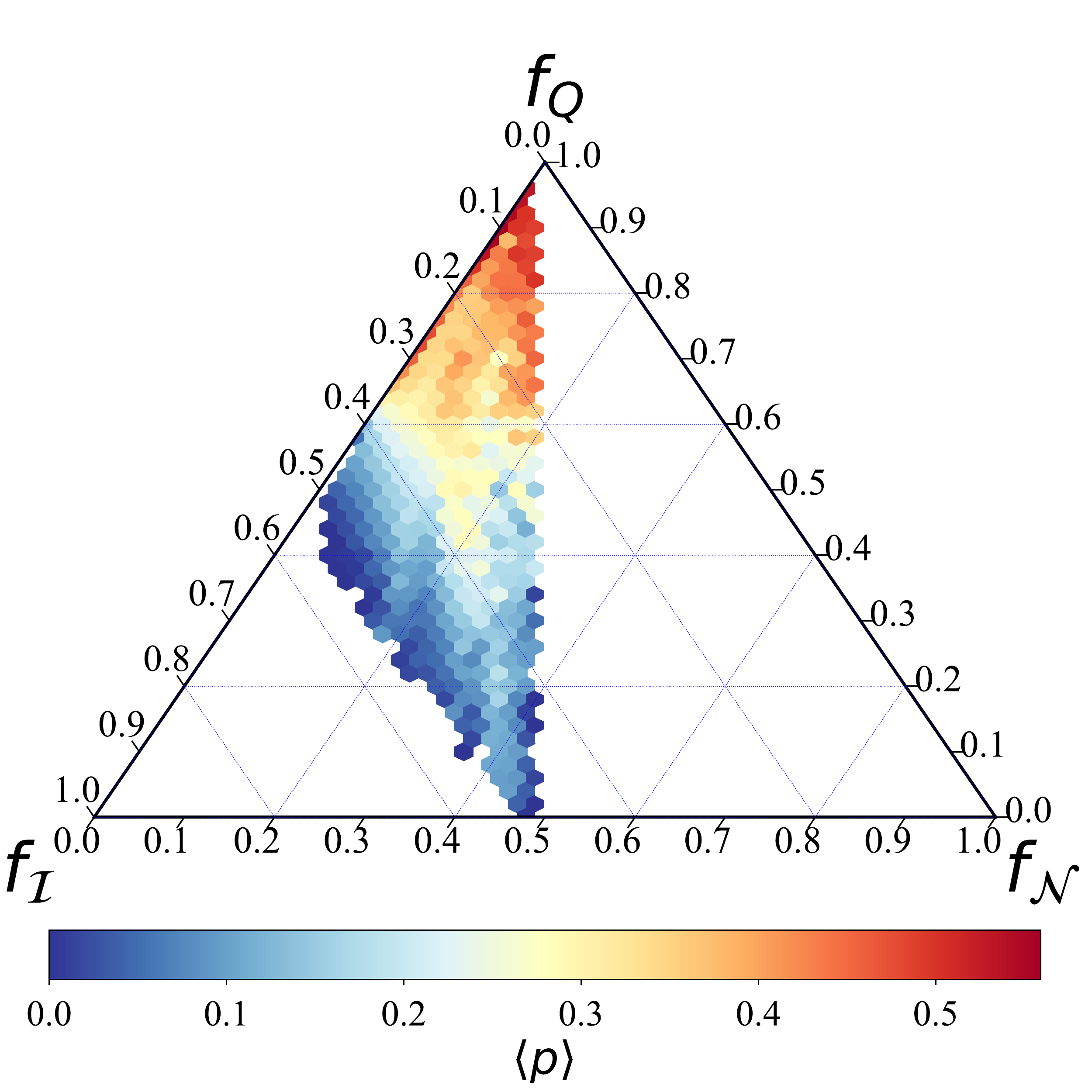}}{0.2in}{.05in}
\topinset{\bfseries(d)}{\includegraphics[width=0.25\columnwidth]{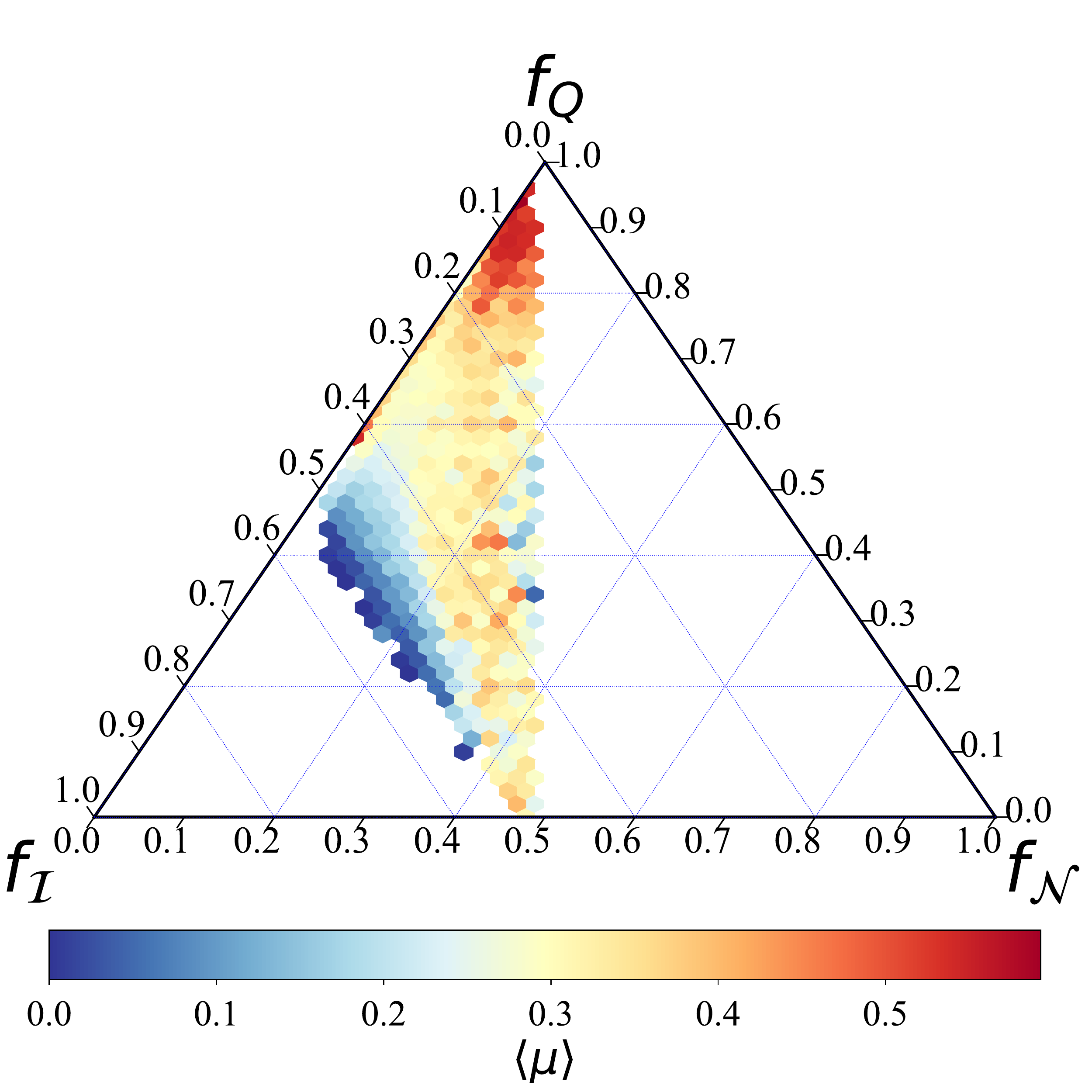}}{0.2in}{.05in}
\end{tabular}
\end{tabularx}
        \caption{Ternary plots representing results for $\sim 2\times10^{5}$ networks as in the main text. In this case, color in each bin of the simplex indicates the average number of blocks $B$ (A); average shape parameter $\xi$ (B); average intra-block noise $p$ (C); and finally average inter-block noise $\mu$ (D).}
        \label{fig:heatmaps_modelVars}
\end{figure}

Figure~\ref{fig:heatmaps_modelVars} shows the results with respect to the parameters of the probabilistic network generation model employed to perform the numerical exploration as explained in the main text (Fig~1). The model is described in detail in the Supplemental Section I. Panel (a)-(d) show the results with respect to varying values of the number of blocks ($B$), shape parameter ($\xi$), intra-block ($p$) and inter-block noise ($\mu$), respectively. The color bar indicates the mean value of the respective parameter in each bin of the simplex.

\section{Supplemental Figure 4}\label{toy_model_heatmaps}
\begin{figure}[H]
\centering
\def\tabularxcolumn#1{m{#1}}
\begin{tabularx}{\linewidth}{@{}cXX@{}}
\def\stackalignment{l}
\begin{tabular}{c}
\def\stackalignment{l}
\topinset{\bfseries(a)}{\includegraphics[width=0.25\columnwidth]{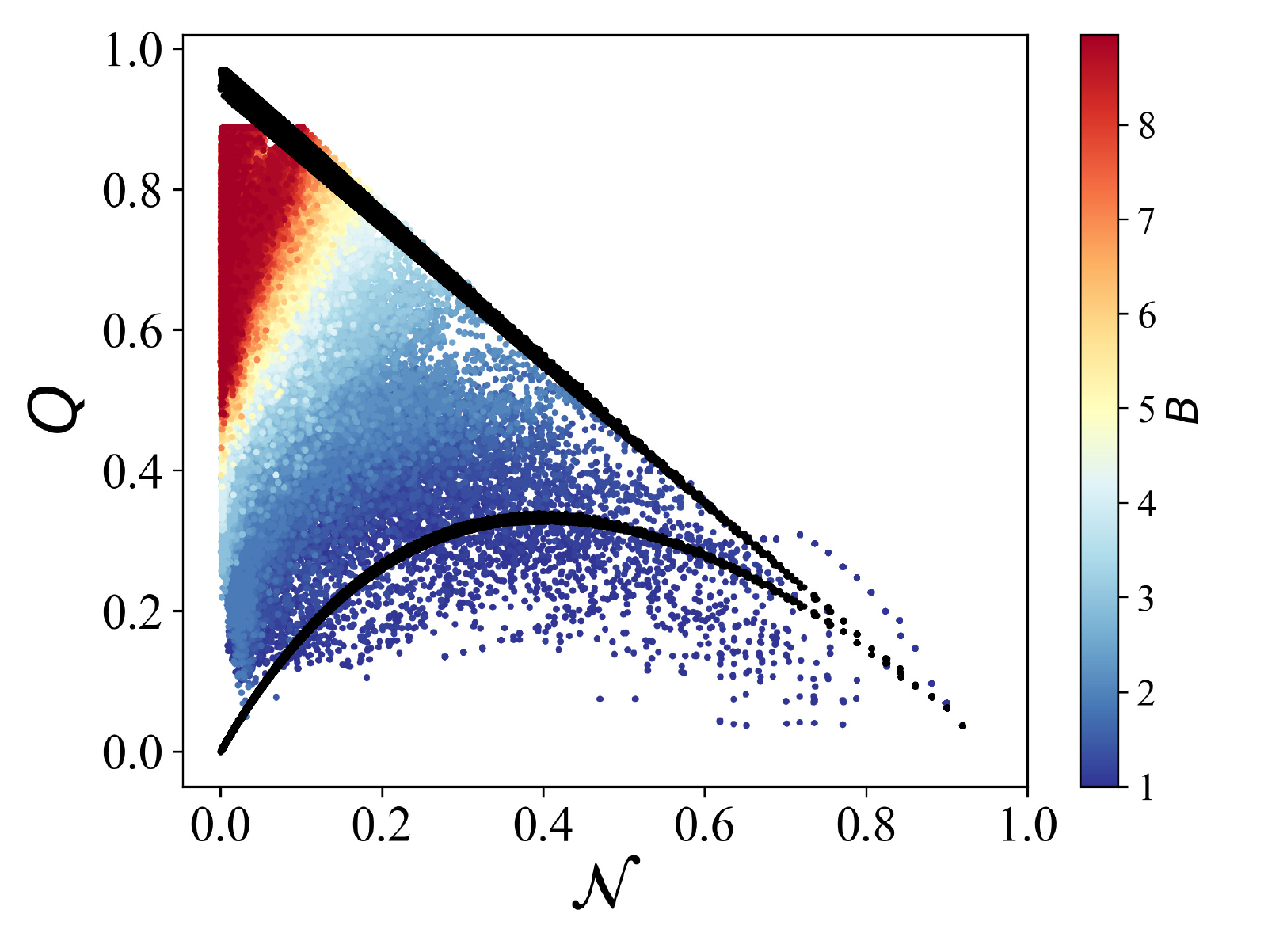}}{0.in}{-0.06in}
\topinset{\bfseries(b)}{\includegraphics[width=0.25\columnwidth]{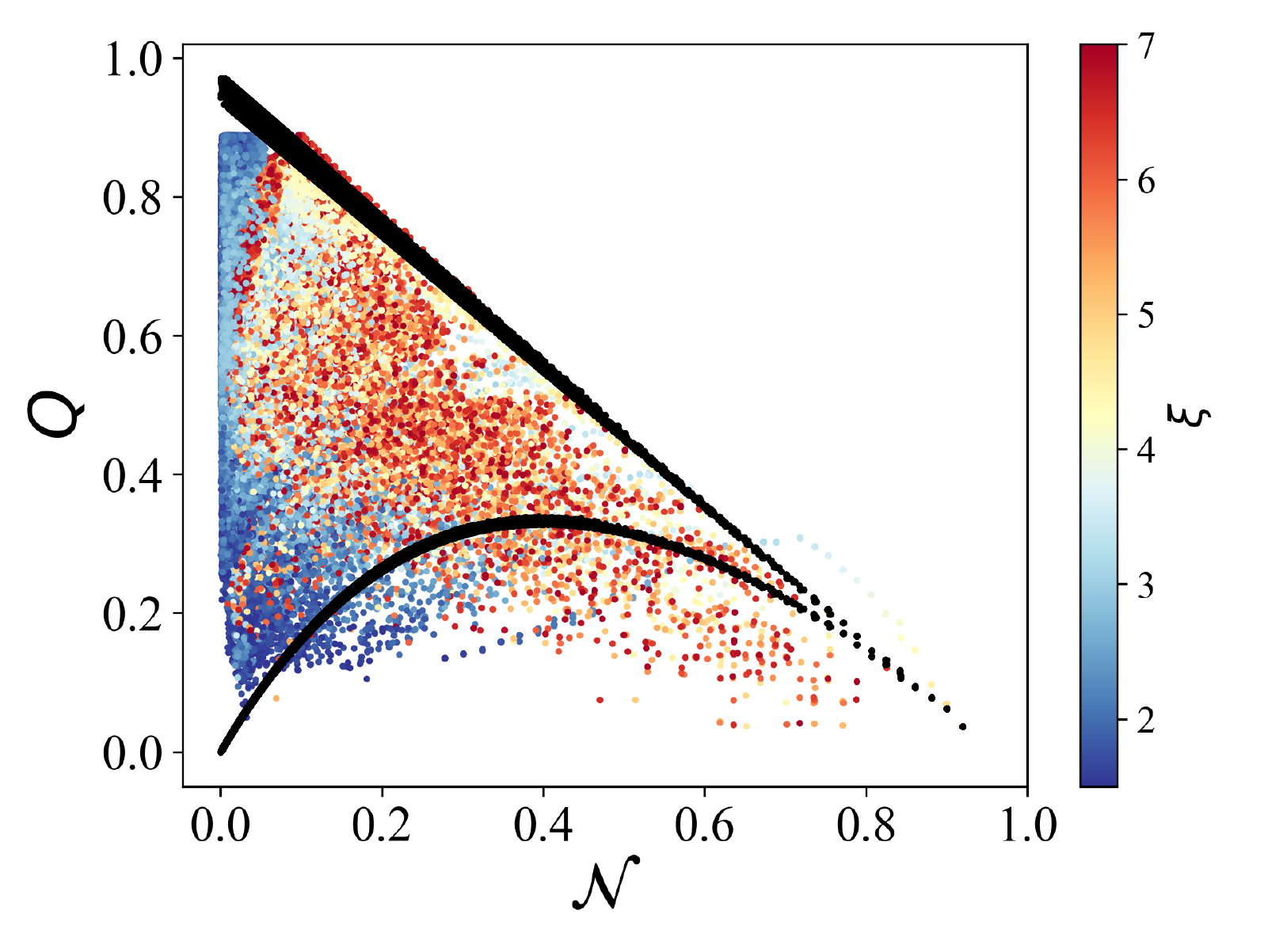}}{0.in}{-.06in}
\topinset{\bfseries(c)}{\includegraphics[width=0.25\columnwidth]{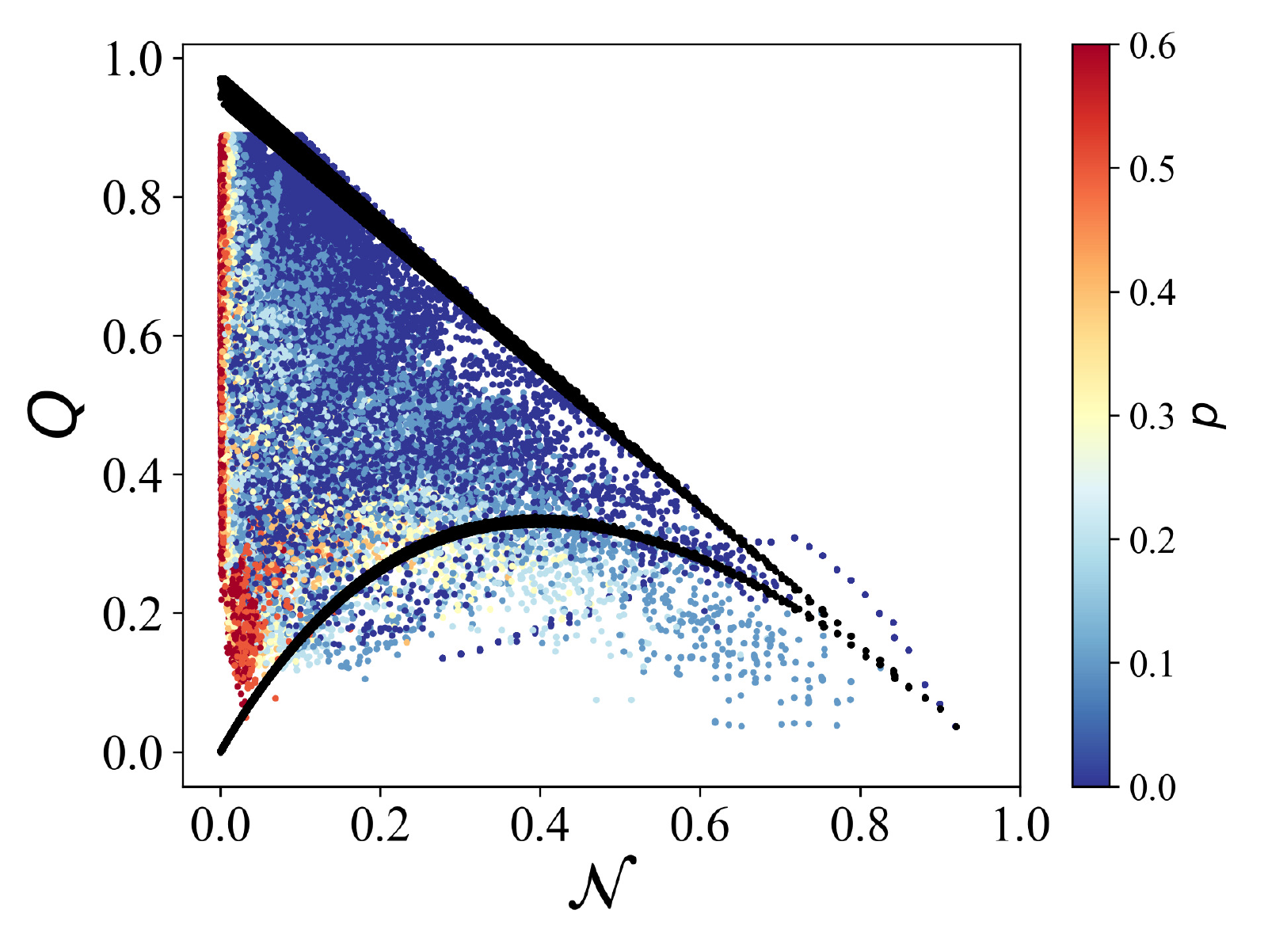}}{0.in}{-.06in}
\topinset{\bfseries(d)}{\includegraphics[width=0.25\columnwidth]{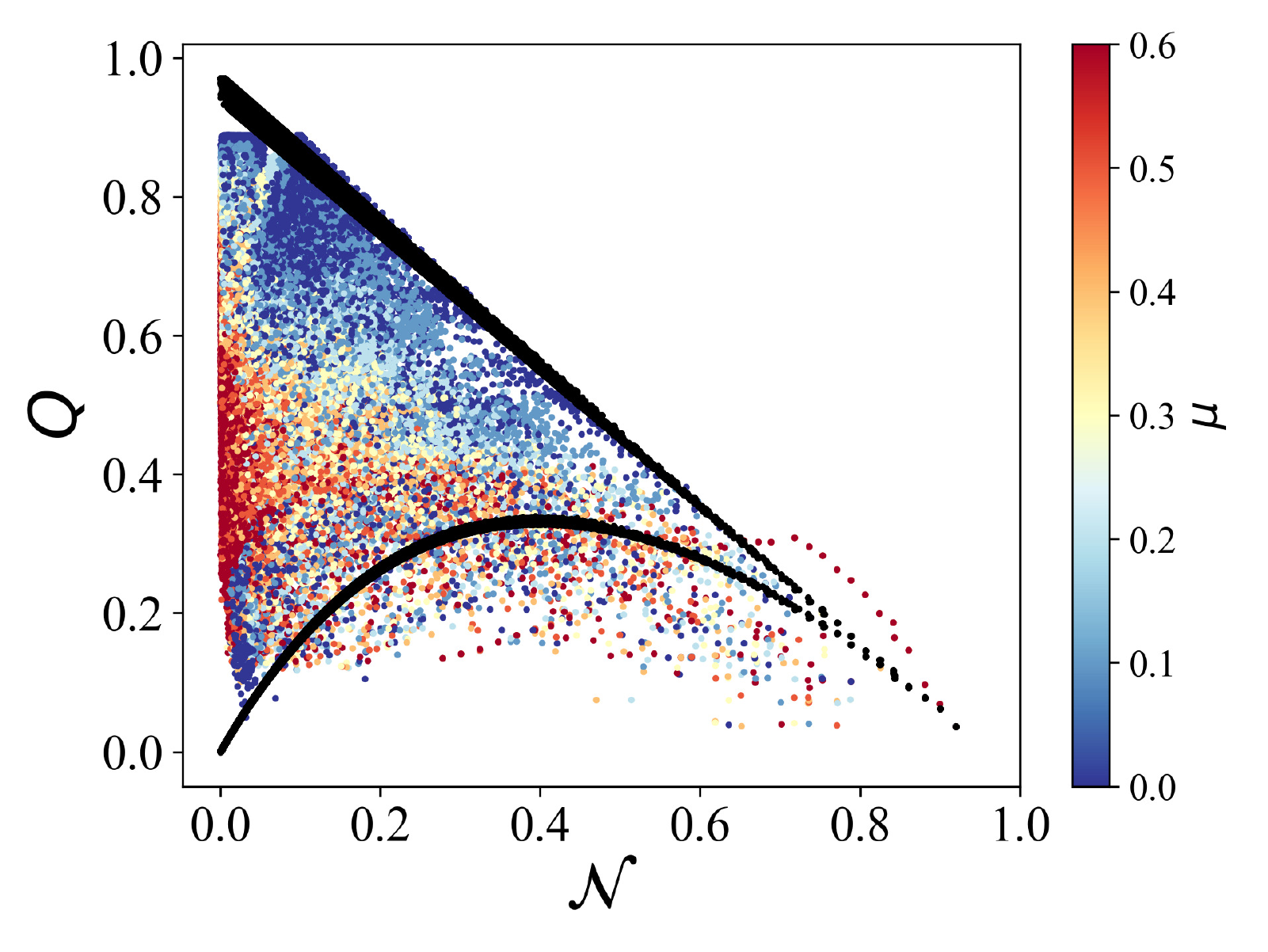}}{0.in}{-.06in}
\end{tabular}
\end{tabularx}
        \caption{Optimized values of $Q$ plotted against $\mathcal{N}$, for the generated networks. The values of the corresponding upper and lower bounds were plotted on top (black dots). The color bar indicates the value of the respective parameters of the probabilistic network generation model (number of blocks ($B$), shape parameter ($\xi$), intra-block ($p$) and inter-block noise ($\mu$), respectively).}
        \label{fig:bounds_modelVars}
\end{figure}

Figure~\ref{fig:bounds_modelVars} shows the values of $Q$ plotted against $\mathcal{N}$, for the all the generated networks employed in the numerical exploration in the main text ($\sim 2\times10^{5}$). The corresponding upper and lower bounds were plotted on top. The color bar, in each case, indicates the values of the respective parameters of the probabilistic network generation model (number of blocks ($B$), shape parameter ($\xi$), intra-block ($p$) and inter-block noise ($\mu$), respectively). 

\section{Supplemental Figures 5 and 6}
\begin{figure}[h]
\includegraphics[width=.4\columnwidth]{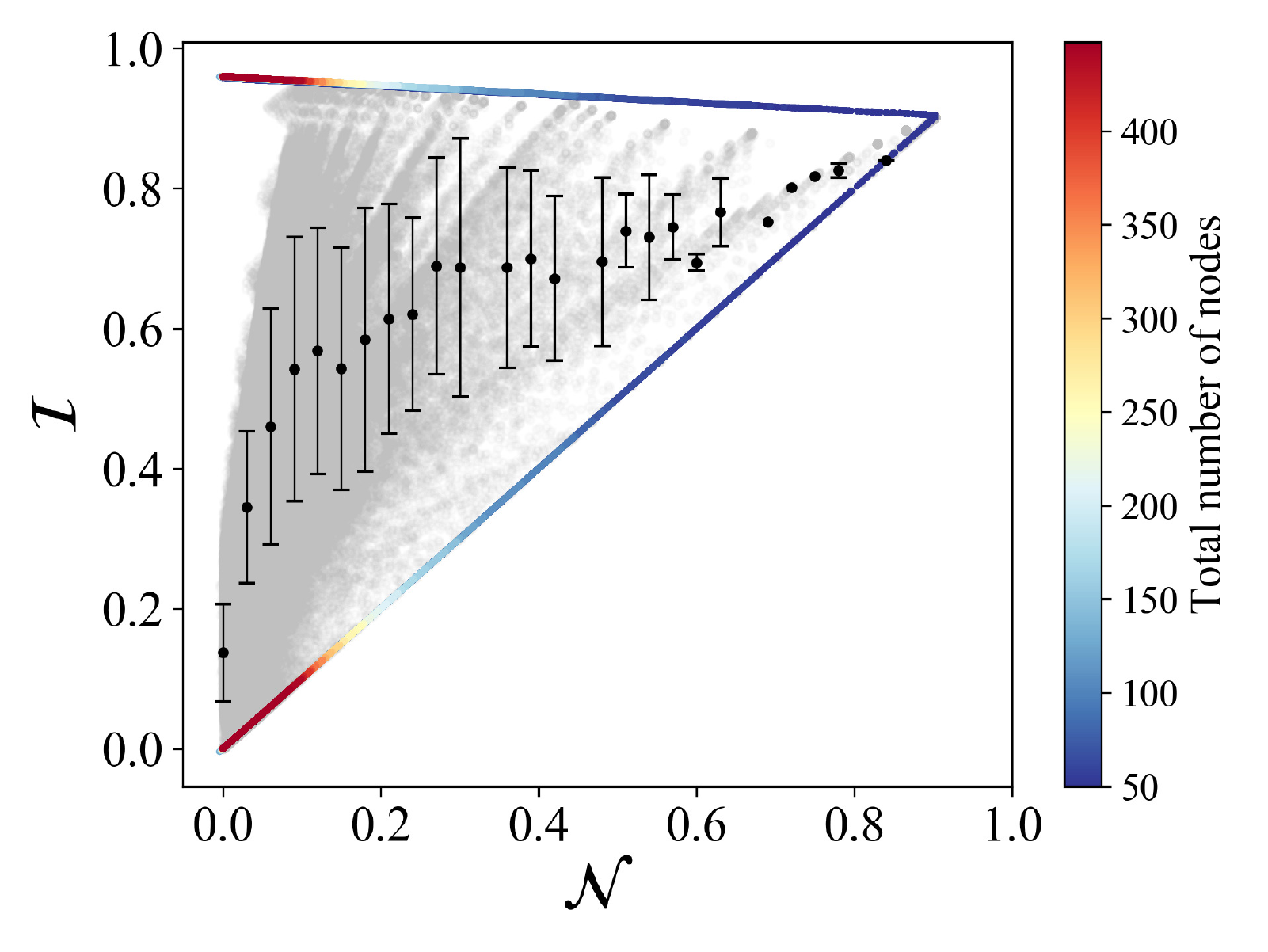}
\caption{Values of $\mathcal{I}$ obtained after optimization (grey dots), plotted against $\mathcal{N}$ for the generated networks. Upper and lower bounds of $\mathcal{I}$ are plotted in colors. The color bar indicates the network's size}
\label{fig:I_vs_N}
\end{figure}

\begin{figure}[H]
\centering
\def\tabularxcolumn#1{m{#1}}
\begin{tabularx}{\linewidth}{@{}cXX@{}}
\def\stackalignment{l}
\begin{tabular}{c}
\def\stackalignment{l}
\topinset{\bfseries(a)}{\includegraphics[width=0.25\columnwidth]{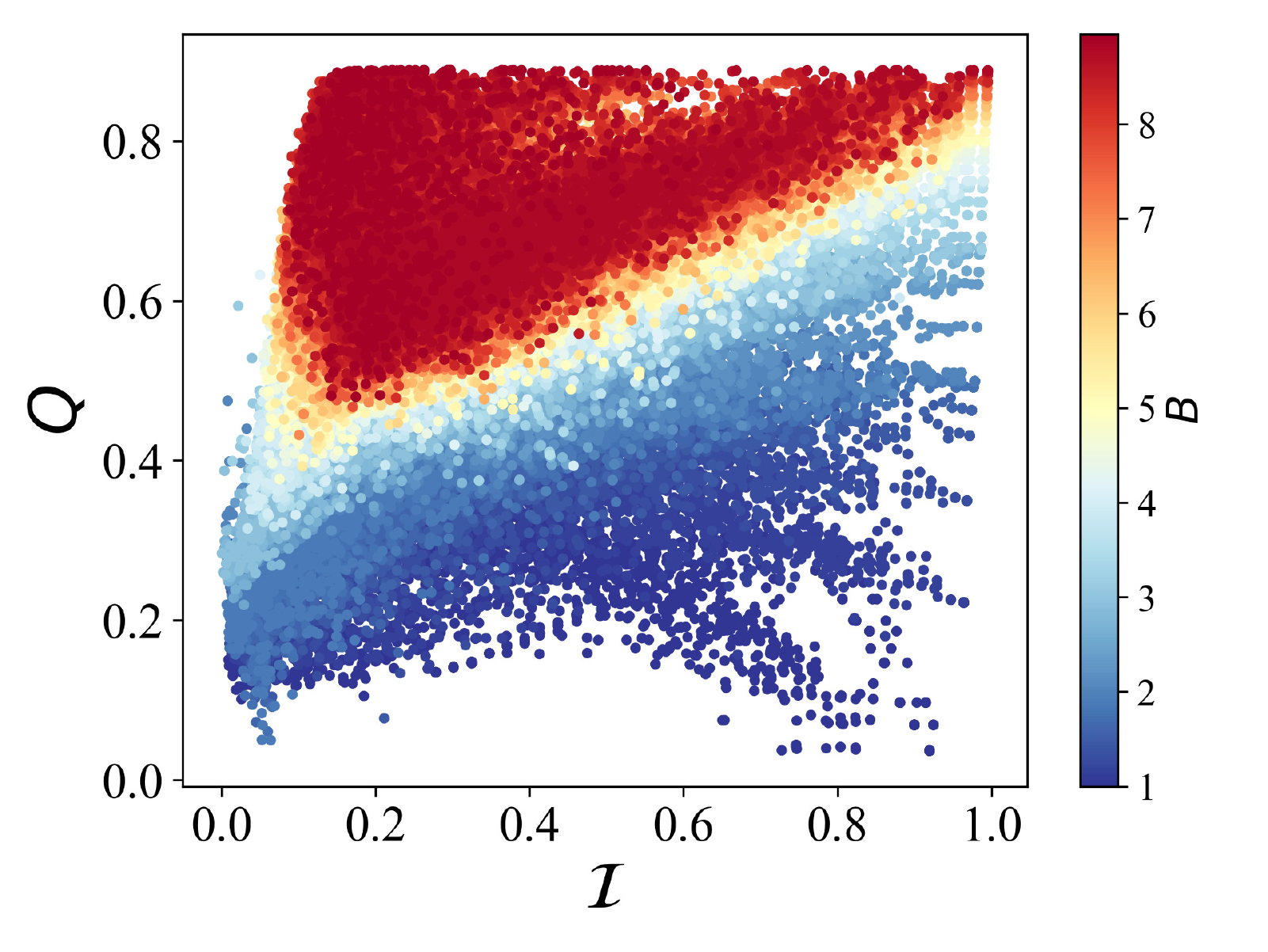}}{0.in}{-0.06in}
\topinset{\bfseries(b)}{\includegraphics[width=0.25\columnwidth]{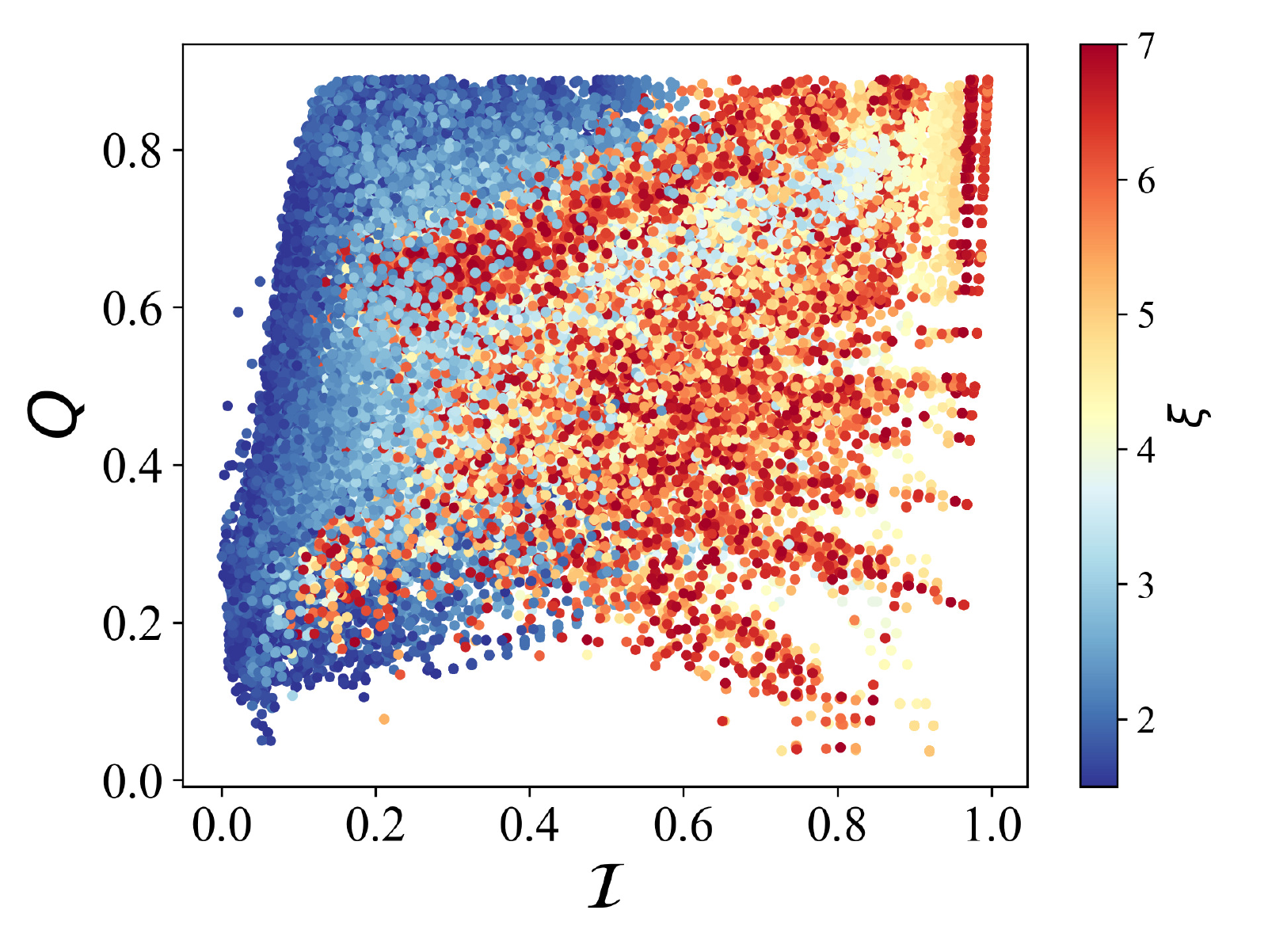}}{0.in}{-.06in}
\topinset{\bfseries(c)}{\includegraphics[width=0.25\columnwidth]{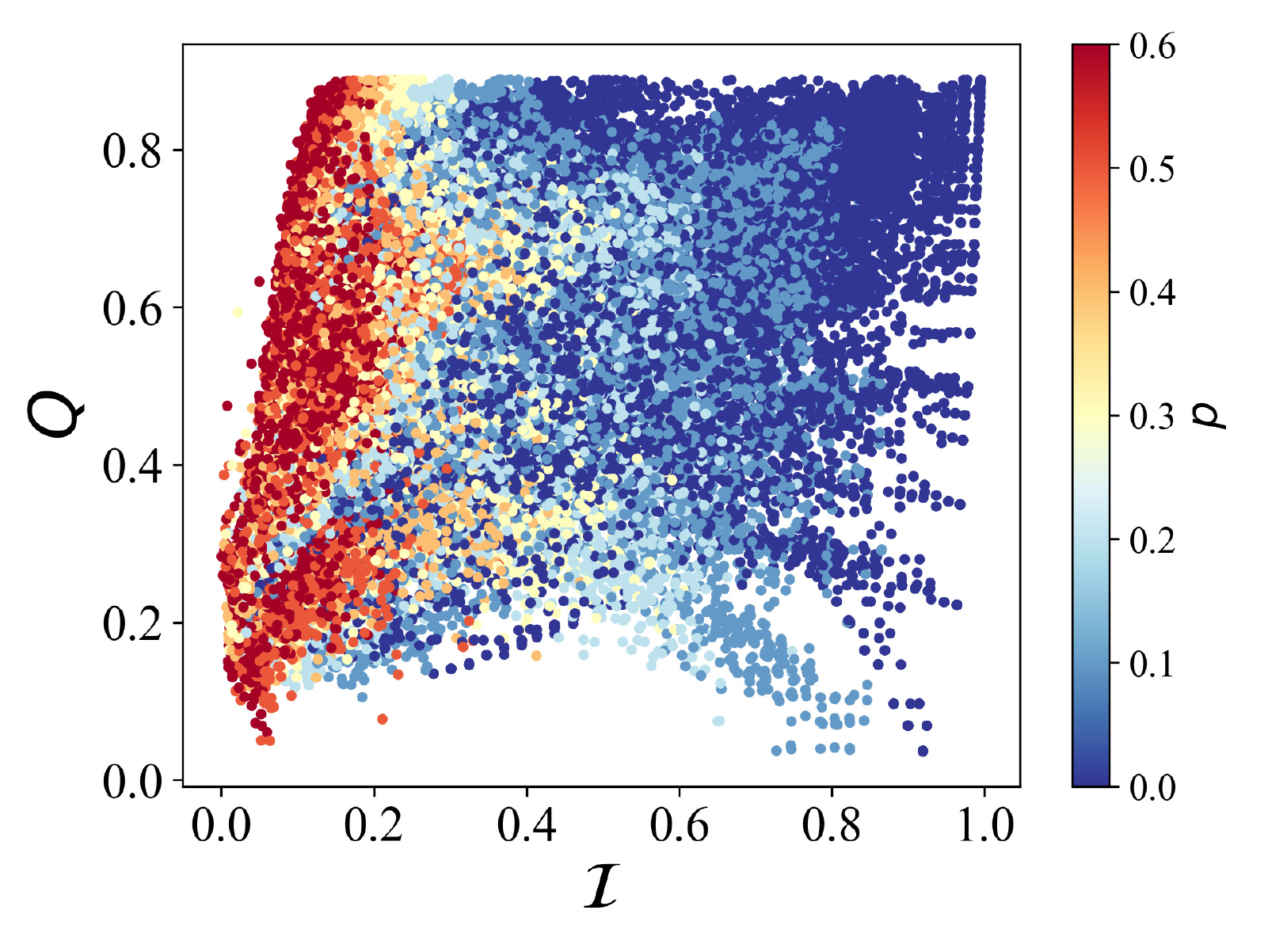}}{0.in}{-.06in}
\topinset{\bfseries(d)}{\includegraphics[width=0.25\columnwidth]{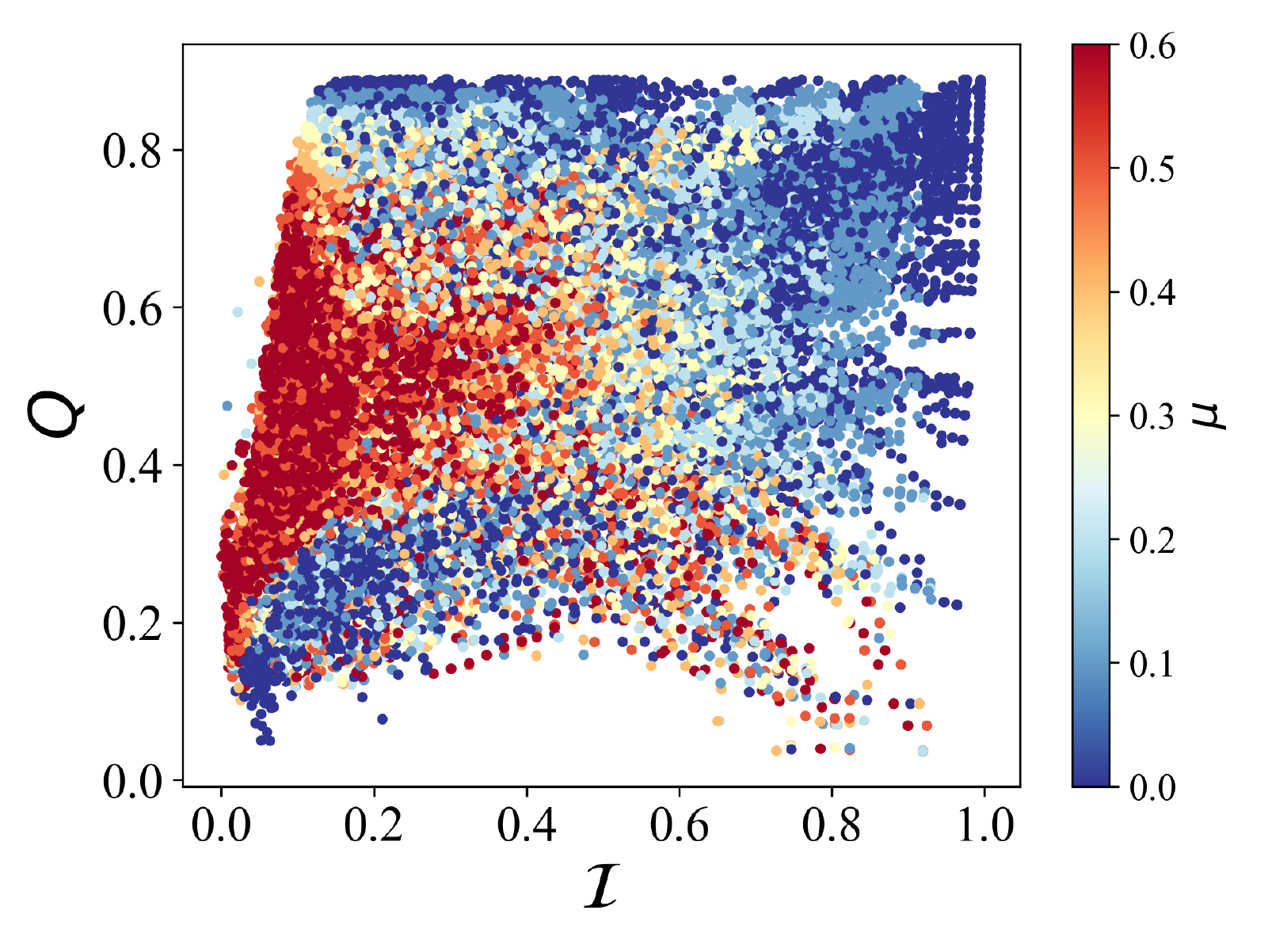}}{0.in}{-.06in}
\end{tabular}
\end{tabularx}
        \caption{Optimized values of $Q$ plotted against the optimized values of $\mathcal{I}$, for the generated networks. The color bar indicates the value of the respective parameters of the probabilistic  network generation model. Panel (a) show the results with respect to the number of blocks. Panel (b) correspond to the  shape parameter $\xi$. Panels (c) and (d) correspond to the noise parameters $p$ and $\mu$, respectively. }
        \label{fig:Q_vs_I}
\end{figure}

For the sake of completeness, we have plotted the values of $\mathcal{I}$ against $\mathcal{N}$ (Fig.~\ref{fig:I_vs_N}) and $Q$ against $\mathcal{I}$ (Fig.~\ref{fig:Q_vs_I}). 

Similar to Fig.~3A in the main text, upper and lower bounds for $\mathcal{I}$ have been calculated in Fig.~\ref{fig:I_vs_N}, taking actual measurements of $\mathcal{N}$ as a starting point. Remarkably, none of the optimized values of $\mathcal{I}$ violates such bounds. This is no surprise with respect to lower bounds, since $\mathcal{I}$ reduces to $\mathcal{N}$ when $B=1$, thus the lower bound simply represents the hard limit $\mathcal{I}=\mathcal{N}$. But even the upper bounds, which represent an estimation, are in excellent agreement with respect to the optimized values of $\mathcal{I}$.

On the other hand, the main lesson from Fig.~\ref{fig:Q_vs_I} is the fact that, unlike $Q$ and $\mathcal{N}$, other patterns can coexist, i.e. there is no clear map between $Q$ and $\mathcal{I}$.
